\documentclass[fleqn,usenatbib]{mnras}
\usepackage{fix-cm}
\usepackage{newtxtext,newtxmath}
\usepackage[T1]{fontenc}
\DeclareRobustCommand{\VAN}[3]{#2}
\let\VANthebibliography\thebibliography
\def\thebibliography{\DeclareRobustCommand{\VAN}[3]{##3}\VANthebibliography}

\usepackage{graphicx}	
\usepackage{amsmath}	
\usepackage{xcolor}	    
\usepackage{physics}    

\newcommand*    \msun{{\,\rm{M}_{\odot}}}

\newcommand*    \tb{T_{\rm b}}

\title[Eccentricity of black hole binaries]{Converging on the eccentricity of massive black hole binaries in galactic mergers}

\author[A. Gualandris et al.]{
\parbox{\textwidth}{
Alessia Gualandris,$^{1}$\thanks{E-mail: a.gualandris@surrey.ac.uk}
Justin Read,$^{1}$
Federica Fastidio,$^{1,2,3}$
Julian Chan,$^{1}$
Thibaut L. François,$^{1}$
Nader Khonji,$^{1}$
and Walter Dehnen$^{4}$
}\\\\
$^{1}$School of Mathematics and Physics, University of Surrey, Guildford, GU2 7XH, UK\\
$^{2}$Dipartimento di Fisica ``G. Occhialini", Universit{\'a} degli Studi di Milano-Bicocca, Piazza della Scienza 3, I-20126 Milano, Italy\\
$^{3}$INFN, Sezione di Milano-Bicocca, Piazza della Scienza 3, I-20126 Milano, Italy\\
$^{4}$Astronomisches Rechen-Institut, Zentrum f{\"u}r Astronomie der Universit{\"a}t Heidelberg, M{\"o}nchhofstra\ss{}e 12-14, 69120, Heidelberg, Germany
}

\date{Accepted 2025 December 21. Received 2025 November 28; in original form 2025 September 2}

\pubyear{\the\year{}}

\begin{document}
\label{firstpage}
\pagerange{\pageref{firstpage}--\pageref{lastpage}}
\maketitle

\begin{abstract}
Binaries of supermassive black holes (MBHBs) represent the primary sources of the gravitational wave background (GWB) detectable by Pulsar Timing Arrays (PTAs). The eccentricity with which binaries form in galactic mergers is the key parameter determining their evolutionary timescale from pairing to coalescence. However, accurately determining the binary eccentricity at formation is difficult in simulations due to stochastic effects. We present a numerical study of the formation and evolution of MBHBs that are potential PTA sources. We simulate mergers of equal-mass galaxies on different initial orbits and follow the dynamics of the MBHBs through the hardening phase. We find that low-resolution simulations are affected by stochasticity due to torques from the stellar distribution acting at pericentre passages. The dispersion in binary eccentricity decreases with increasing central resolution, as expected for a Poisson process. We provide a fitting formula for the resolution requirement of an $N$-body simulation of MBHB formation and evolution as a function of the initial eccentricity of the merger, $e_0$, and the required accuracy in the binary eccentricity, $e_{\mathrm{b}}$. We find that binaries experience a torque at first pericentre that is approximately independent of initial eccentricity, producing a general trend in which the binary eccentricity decreases above sufficiently large initial orbital eccentricities. While this behaviour is generic, the precise cross-over eccentricity ($e_0 \sim 0.97$ in our models) and the sharpness of the drop-off depend on the galaxy initial conditions. We provide a fitting formula for $e_{\mathrm{b}}(e_0)$ that can be used in semi-analytical models to determine the merger timescales of MBHBs as well as the amplitude and slope of the GWB.
\end{abstract}




\section{Introduction}

Massive black hole binaries (MBHBs) are expected to form abundantly in the Universe as a result of the hierarchical assembly of structures in the Lambda Cold Dark Matter cosmological model, given the ubiquitous presence of massive black holes (MBHs) at the centre of galaxies \citep{Begelman1980, Roos1981, MV1992}. Their formation and evolution are linked to those of the host galaxies they inhabit, as evidenced by scaling relations \citep[e.g.][]{FerrareseMerritt2000, MarconiHunt2003, KormendyHo2013, ReinesVolonteri2015}, and lead to coalescence by emission of gravitational waves (GWs) at separations between MBHs of order $(10^{-3}-10^{-1})\,\rm pc$, depending on mass. MBHBs represent the loudest GW sources in the Universe; their inspiral signals can be detected by current ground-based detectors such as Pulsar Timing Arrays (PTAs) \citep{EPTA2023search} in the frequency range $(10^{-9}-10^{-7})$ Hz and future space-based detectors such as the Laser Interferometer Space Antenna (LISA) \citep{LISA2017} in the frequency range $(10^{-5}-10^{-1})$ Hz. 

PTAs have recently found evidence for a common red signal in the nano-Hz frequency range, with the quadrupolar inter-pulsar correlation properties of a gravitational wave background (GWB) \citep{EPTA2023search, NANOGRAV2023, Parks2023, Xu2023}. The stochastic, low-frequency signal is consistent with expectations from the population of the most massive MBHBs, but could also arise or contain contributions from a variety of early Universe sources \citep{EPTA2023interp}. 
Among astrophysical MBHBs, the sources that most contribute to the GWB are the binaries with masses in excess of $10^8\msun$ formed in major mergers (mass ratio $q>0.25$) at redshift $z<2$ \citep{Sesana2008}. These are generally found in massive galaxies, with halo and stellar masses $\sim 10^{12.5}\msun$ and $\sim 10^{11}\msun$, respectively \citep{IzquierdoVillalba2023}.

PTAs monitor arrays of millisecond pulsars, rapidly spinning neutron stars emitting extremely regular pulses with periods of order a millisecond, and exploit their remarkable rotational stability to infer the existence of perturbations due to a GWB.
Such a background modulates pulsar arrival times giving rise to a stochastic low-frequency signal, common to all pulsars, with a signature spatial correlation.  The signal can be modelled as a single power-law spectrum 
in the strain, where the spectral index depends on the dominant source of the GWB while the amplitude depends on the properties of each source. 
The slope of the GWB is not strongly constrained by current PTA data; however there is evidence for a flattening of the spectrum at the lowest frequencies, where the data is most informative. This departure from the $\alpha=-2/3$ value expected for a population of circular binaries evolving solely due to GW emission suggests that binaries may not be circular and that eccentricity and coupling with the local environment may be important \citep{EPTA2023interp}. 

Galactic mergers are expected to 
occur on highly radial or parabolic orbits \citep{Benson2005, KhochfarBurkert2006, Wetzel2011, Jiang2015}. Recently, \citet{Fastidio2024} studied the properties of major galactic mergers in the IllustrisTNG100-1 cosmological simulation \citep{Pillepich2018}. They found that the majority of galactic mergers with mass ratio $q > 0.1$ at redshift $z < 2$ occur on nearly radial orbits, with a mean eccentricity of 0.88, a median eccentricity of 0.97, and with approximately 60\% of them having eccentricity $0.95 \leq e < 1$. The eccentricity distribution upon entry in the PTA frequency band spans the range (0.2--0.8), which is sufficient to account for the observed flattening in the GWB at low frequencies. Furthermore, their study of consecutive low-redshift major mergers with initial conditions drawn from IllustrisTNG100-1 reveals a set of 8 MBHBs that enter the PTA band with $e>0.85$ \citep{Fastidio2025}. A high eccentricity will also result in shorter evolutionary timescales, affecting the amplitude as well as the slope of the signal.

The eccentricity of MBHBs upon entrance in the PTA frequency range $(10^{-9}-10^{-7})$ Hz depends on the properties of the binaries at formation and their evolution through the hardening phase. The latter is relatively well understood based on the results of scattering experiments \citep[e.g.][]{Quinlan1996, Sesana2006,Sesana2015} and $N$-body simulations \citep[e.g.][]{GualandrisMerritt2012, Rantala2017, Khan2018, Mannerkoski2019, Rawlings2025}. These show a tendency for the eccentricity to increase due to scatterings with stars, unless the MBHB is almost circular, in which case it tends to remain circular. The eccentricity of MBHBs at formation, however, is challenging to constrain due to numerical stochasticity: \citet{Nasim2020} show a significant spread in eccentricity at the time of MBH binding among a set of simulations of eccentric ($e_0=0.9$) mergers involving different random realisations of identical galaxy models.
Such stochasticity is found to decrease as $1/\sqrt{N}$, where $N$ is the total particle number in the simulation, taken as a measure of resolution. This is interpreted as evidence of Poisson-type noise that affects the properties of the MBHs at the time of binding. \citet{attard2024} perform simulations of mergers with moderately eccentric orbits ($e_0=0.7$) and find a similar behaviour: the scatter in binary eccentricity is always larger than that in semi-major
axis and scales with resolution. Interestingly, \citet{Gualandris2022} report an identity relation between the eccentricity of MBHBs at the time of binary formation and the initial eccentricity of the merger, sampled up to $e_0=0.9$. Such relation persists to later times, albeit with scatter, opening up the possibility to implement this scaling in semi-analytical models. However, \citet{Fastidio2024} show that the relation breaks, at least in simulations with modest resolution, for almost radial orbits. In this case, binaries tend to form with a lower eccentricity than expected based on the relation of \citet{Gualandris2022} and show a much larger spread. The same breaking and increased scatter is observed in the simulations of \citet{Rawlings2023}. However, they attribute the behaviour to the nature of nearly radial encounters.

In this work, we investigate the origin of the stochasticity in the eccentricity of MBHBs at formation and the apparent distinct behaviour of almost radial ($e_0=0.99$) galactic mergers. We perform $N$-body simulations of MBHBs formation and evolution in equal mass mergers of moderately cuspy gas free galaxies using the FMM code {\tt Griffin}. Our study focuses on MBHs in the range $10^7 - 10^{10}\msun$, corresponding to the population probed by PTAs. We find that the dispersion in the eccentricity at binary formation scales with resolution as a Poisson process, for any value of the initial eccentricity of the galactic orbit up to $e_0=0.99$, but is always larger for more eccentric orbits. We confirm that the identity relation of \citet{Gualandris2022} between $e_{\mathrm{b}}$ and $e_0$ breaks above a critical eccentricity: in this case binaries form with a lower eccentricity. The turnover in the eccentricity distribution owes to the effect of global torques from the stellar distribution, acting primarly at pericentre passages.  These induce changes in angular momentum that are increasingly more impactful the more radial the initial orbit is. Above a critical merger eccentricity that depends on the galactic model, binaries emerge from the first and second pericentre passages with an increased angular momentum and therefore a lower eccentricity.

\section{Methodology}
In order to explore the nature and origin of stochasticity in the eccentricity of MBHBs at formation, we perform a resolution study of galactic mergers, varying the initial orbit of the interacting galaxies. We consider single-component galaxy models representing stellar bulges, as in \citet{Nasim2020}, following a density profile of the type \citep{Dehnen1993}
\begin{equation}
\label{eqn:dehnen_density}
	\rho(r) = \frac{(3-\gamma)M}{4\pi} \frac{r_0}{r^{\gamma} (r+r_0)^{4-\gamma}}\,,
\end{equation} 
where $M$ is the total galaxy mass, $r_0$ is the scale radius and $\gamma$ is the inner slope. Each galaxy follows a shallow profile with slope $\gamma=0.5$ and hosts a central MBH of mass $M_{\bullet} = 0.0025\,M$. We adopt normalised units such that $M_{\rm tot}=G=r_0=1$, where $M_{\rm tot}$ is the total mass of the merging galaxies. Scaling to physical units can be performed by equating the value of the central MBH's influence radius measured in the initial model according to
\begin{equation}
M(<R_{\rm infl}) = 2 M_{\bullet}
\end{equation}
with the value obtained from the relation
\begin{equation}
    R_{\rm infl} = \frac{GM_{\bullet}}{\sigma^2}
\end{equation}
where $\sigma$ is the stellar velocity dispersion obtained from the $M-\sigma$ relation of \citet{KormendyHo2013}
\begin{equation}
    \log_{10} \left( \frac{M_{\bullet}}{\msun} \right)
  = 8.2 + 4.3 \, \log_{10} \left( \frac{\sigma}{200~\mathrm{km~s^{-1}}} \right).
\end{equation}

Length, time and velocity units are given in Table \ref{tab:units} for a set of MBH masses representative of PTA sources (see Section \ref{sec:times}).

\begin{table}
\begin{center}
\label{tab:units}
\begin{tabular}{lcccc} 
\hline 
$M_{\bullet} (\msun)$ & $\sigma ($km s$^-1$) & [L] (pc) & [T] (yr) & [V] (km s$^-1$)\\
\hline
$1\times10^{7}$ & 105.2 &  3.9 & $3.6\times10^{4}$ & 105.2 \\
$1\times10^{8}$ & 179.7 & 13.3 & $7.3\times10^{4}$ & 179.6 \\
$1\times10^{9}$ & 307.0 & 45.7 & $1.5\times10^{5}$ & 306.9 \\
$1\times10^{10}$ & 524.4 & 15.6 & $2.9\times10^{5}$ & 524.2 \\
\hline
\end{tabular} 
\caption{Scaling to physical units [L] (distance), [T] (time) and [V] (velocity) for different adopted values of the binary mass $M_{\bullet}$ and stellar velocity dispersion $\sigma$.}
\end{center}
\end{table}

We model equal mass mergers of galaxies on bound Keplerian orbits with initial eccentricity $e_0=0.9, 0.97, 0.99$.  Large eccentricities are expected for galactic mergers in a cosmological context, as shown e.g. in \citet{KhochfarBurkert2006,Fastidio2024} for the sample of major mergers which are hosts to PTA sources in Illustris-TNG100 \citep{Pillepich2018}.

The progenitors are initially placed in the $x-y$ plane at a distance $d = 20 \,r_0$ along the $x$-axis. 
For each combination of parameters, we 
vary the total particle number from $N = 1\mathrm{M}$ to $128\mathrm{M}$ (with $1\mathrm{M}=2^{20}=1048576$) and,  for each resolution, ran $N_{\mathrm{sim}}$ simulations started from different random realisation, see Table \ref{tab:init}.

We measure the resolution in each model in two ways: (i) as the MBH-to-star mass ratio $M_{\bullet}/m_*$ and (ii) as the number of stars enclosed within 5 radii of influence of the primary MBH at the initial time $N(< 5 R_{\rm infl})$. 
The latter metric can be applied to any stellar model, regardless of the chosen density profile, and can therefore aid comparisons with other simulations.

\begin{table}
\begin{tabular}{rcccc}
\hline 
\multicolumn{1}{c}{$N$} & $e_0$ & $M_{\bullet}/m_*$ &  N(< 5 $R_{\rm infl})$ & $N_{\mathrm{sim}}$ \\
\hline 
1\,M & 0.9, 0.97, 0.99 & $3.81\times10^{-4}$  & $8.65\times10^4$ & 8\\
2\,M & 0.9, 0.97, 0.99 & $1.91\times10^{-4}$  & $1.73\times10^5$ & 8\\
4\,M & 0.9, 0.97, 0.99 & $9.54\times10^{-5}$  & $3.43\times10^5$ & 8\\
8\,M & 0.9, 0.97, 0.99 & $4.77\times10^{-5}$  & $6.85\times10^5$ & 8\\
16\,M & 0.9, 0.97, 0.99 & $2.38\times10^{-5}$  & $1.38\times10^6$ & 8\\
32\,M & 0.9, 0.97, 0.99 & $1.19\times10^{-5}$  & $2.74\times10^6$ & 8\\
64\,M & 0.9, 0.97, 0.99 & $5.96\times10^{-6}$  & $5.50\times10^6$ & 3\\
128\,M & 0.9, 0.97, 0.99 & $2.98\times10^{-6}$  & $1.10\times10^7$ & 1\\
\hline
\end{tabular}
\caption{Initial parameters of the galaxy mergers. From left to right: Total particle number in the merger, with $1\,\mathrm{M}=1048576$ particles; initial eccentricity of the galactic merger; MBH to stars mass ratio; number of stars enclosed within 5 radii of influence of the primary MBH; number of simulations, each using another random realisation.} 
\label{tab:init}
\end{table}

We evolve all models with the code {\tt Griffin}, an adaptive Fast Multipole Method (FMM) $N$-body solver with an $\mathcal{O}(N)$ scaling that can achieve the accuracy of direct-summation codes \citep{Dehnen2014}. This is obtained with a careful choice of parameters, informed by an estimate of the actual force error incurred by individual interactions. All star-star interactions are modelled with FMM, whereas all interactions involving a MBH are integrated with direct summation. We assign a softening length  $\epsilon=0.023$ to star-star interactions and $\epsilon_\bullet = 0.01$ to MBH-MBH and MBH-star interactions. We caution that a few pericentre passages may reach distances slightly below $\epsilon_\bullet $ for the most eccentric binaries. However, these happen well after binary formation and do not affect the analysis of binaries at formation.

The models are followed with {\tt Griffin} through the galactic merger and the formation of the binary and into the hardening phase. The simulations are halted at time $t=200$ in normalised units, well into the collissionless loss cone refilling regime.

\section{Results}
\begin{figure*}
	\includegraphics[width=1.04\columnwidth]{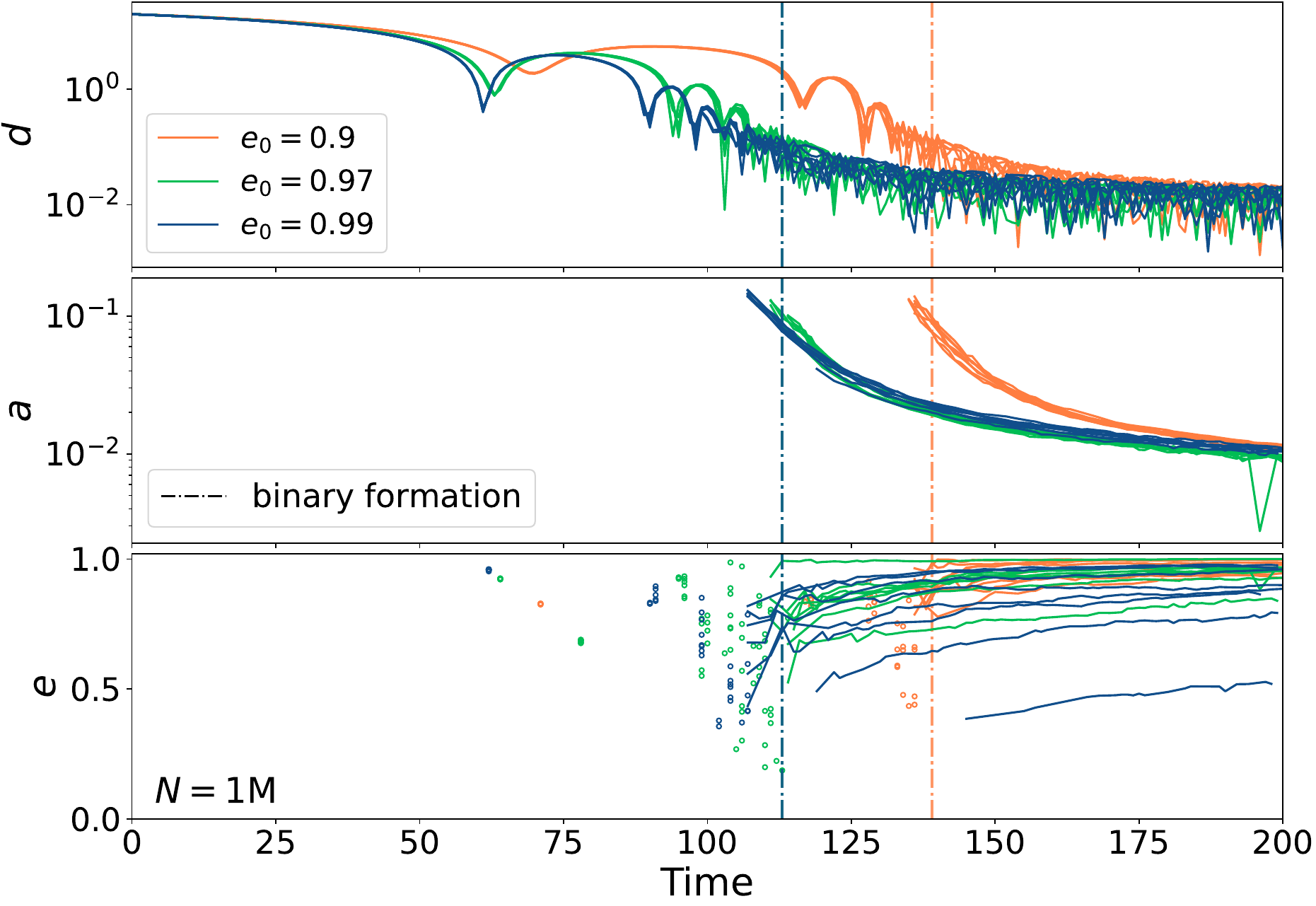}\hfill
    \includegraphics[width=1.04\columnwidth]{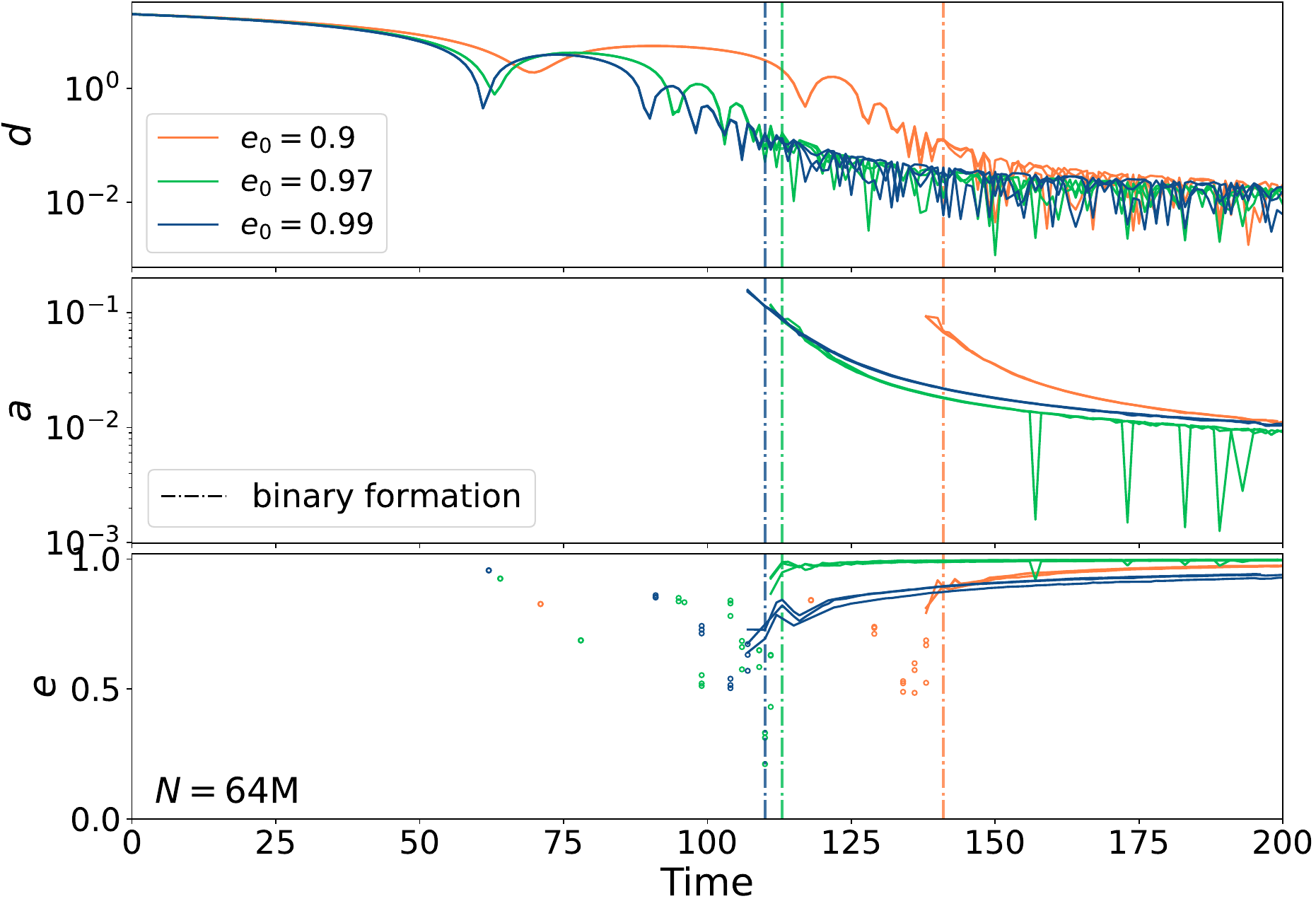}
    \vspace*{-5mm}
    \caption{Distance between the MBHs, semi-major axis and eccentricity of the binary pairs as a function of time for the different values $e_0$ of the initial eccentricity of the galactic orbit. The orbital elements (lines) are computed assuming a Keplerian isolated 2-body system. The left hand side shows models at low resolution, with total particle number $N=1\mathrm{M}$, while the right hand side shows models at high resolution, with $N=64\mathrm{M}$. The different coloured lines and points represent the different random realisations for each value of the initial eccentricity $e_0$. The bottom panel shows the Keplerian eccentricity (lines) and the orbital eccentricity (points) calculated directly from the pericentre and apocentre values of the orbit, at each pericentre passage from the start of the simulation to pairing. The vertical lines show the time of binary formation $\tb$ for one random realisation of each $N$ and $e_0$.}
    \label{fig:orb_elements}
\end{figure*}

We compute the distance between the MBHs and the Keplerian 2-body orbital elements as a function of time, for all models. These are shown in Fig.~\ref{fig:orb_elements} for the low ($N=1\mathrm{M}$) and high ($N=64\mathrm{M}$) resolution models, for the three different values of the orbital eccentricity $e_0=0.9, 0.97, 0.99$. Vertical lines mark the time of binary formation for one random realisation at a given $N$ and $e_0$. Circles indicate the orbital eccentricity of the MBH pair computed directly from the pericentre and apocentre of the orbits, and plotted at pericentre passages. These are particularly useful to determine the properties of the forming binaries during the merger, i.e. while the MBHs are still unbound.
We see that higher orbital eccentricities for the galactic merger result in shorter times to pericentre as well as deeper pericentre passages in the galactic potential. Interestingly, orbital eccentricities of the pairs match the initial eccentricity $e_0$ of the galactic orbit at the first pericentre passage. They then decrease due to the circularising effect of dynamical friction, followed by a rapid increase prior to binding. 
Stochastic effects are apparent in all cases: we discuss their implications in the following subsections.

\subsection{Numerical stochasticity}
Numerical stochasticity refers to the spread in the orbital elements of the MBHB at (and after) formation and has been reported in the literature \citep{Nasim2020, Rawlings2023, attard2024}. Consistent with their results, we find that the scatter in eccentricity is always larger than that in semi-major axis, given the larger susceptibility of angular momentum (and therefore eccentricity) to perturbations. As is appreciable from Fig.~\ref{fig:orb_elements}, both appear to decrease with increasing particle number, though convergence in eccentricity is slower. 

\begin{figure}
 \includegraphics[width=\columnwidth]{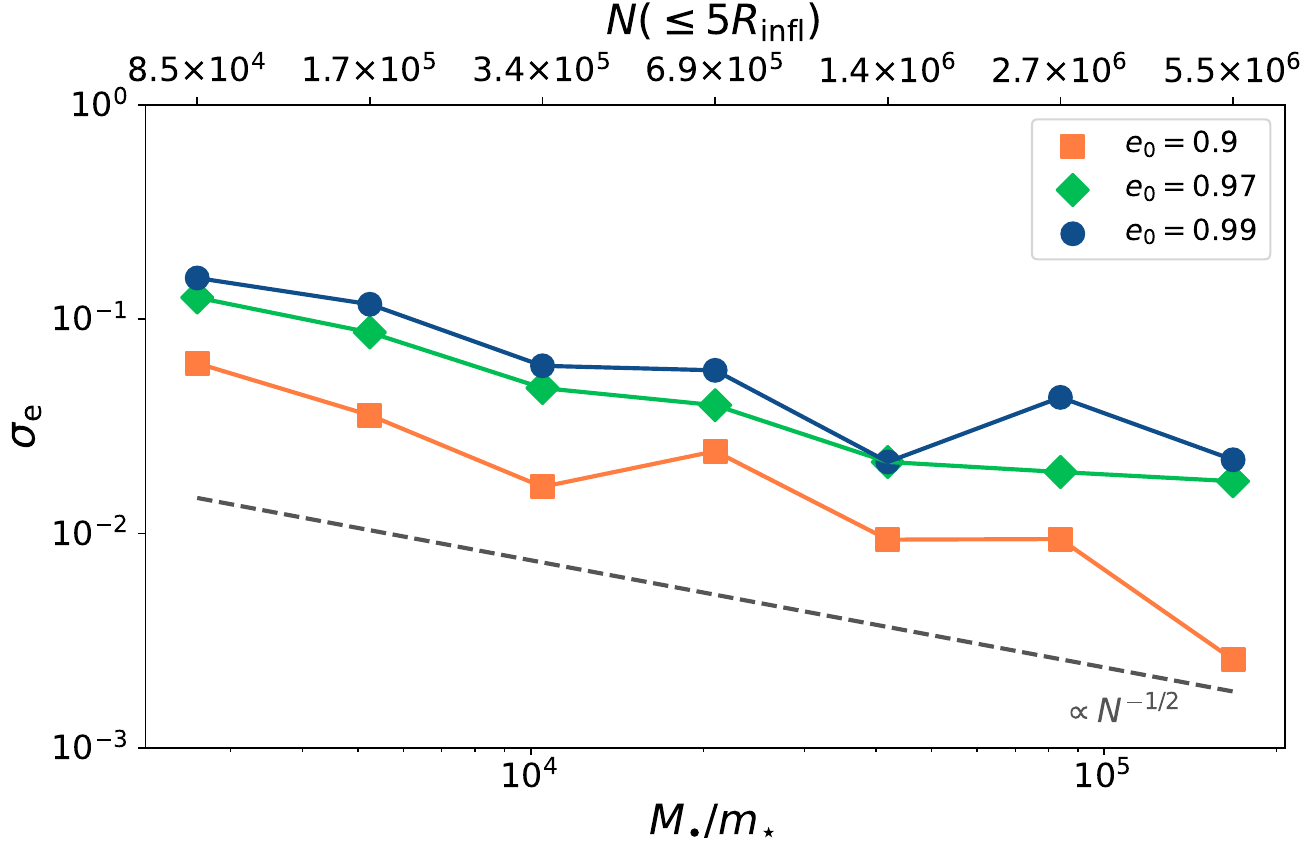}
    \vspace*{-5mm}
    \caption{Dispersion in eccentricity as a function of mass resolution and the number of stars enclosed within 5 influence radii of the binary. The eccentricity is taken at the time of binary formation.  The dispersion is larger in the models with $e_0=0.99$ at all resolutions, but follows the same $\sim 1/\sqrt{N}$ scaling as the dispersion in the $e_0=0.9$ models.}
    \label{fig:disp_ecc}
\end{figure}

To quantify the amount of numerical stochasticity and how this scales with the parameters of the models, we compute the dispersion in eccentricity at the time of binary formation over the different random realisations available for each combination of model parameters, as detailed in Table ~\ref{tab:init}. We define the time $\tb$ of binary formation as the time when the binding energy of the binary pair becomes negative and remains negative, thereby avoiding the chaotic transitionary phase described in \citet{Gualandris2022}, after having removed pericentre passages from the data. We find $\tb=141, 113, 110$ for $e_0 = 0.9,0.97, 0.99$, respectively. We show the dispersion in eccentricity at time $\tb$ in Fig.~\ref{fig:disp_ecc} as a function of resolution, measured in two ways: (i) as the MBH to star mass ratio $M_{\bullet}/m_*$ and (ii) as the number of stars enclosed within 5 radii of influence of a single MBH at the initial time, $N(< 5 R_{\rm infl})$. The radius is computed in the isolated progenitor galaxy at time $t=0$: $R_{\rm infl}\sim 0.19$. The dispersion decreases with increasing resolution, roughly following the $\sim1/\sqrt{N}$ scaling characteristic of a Poisson process. This holds true for any value of the initial orbital eccentricity $e_0$ of the merger, though the dispersion at a fixed resolution is larger for more radial orbits, and in particular for the $e_0=0.99$ models. This is qualitatively consistent with prior work: \citet{Rawlings2023} compare random realisations of models at $e_0=0.9$ and $e_0=0.99$ and find a larger dispersion in the almost radial orbits; \citet{Fastidio2024} consider individual mergers with cosmological initial conditions and find an increased spread for orbits with $e_0>0.9$. 

We interpret this trend as evidence that the dispersion in eccentricity is due to limited resolution rather than an intrinsic characteristic of the pairing process. 

\begin{figure}
	\includegraphics[width=\columnwidth]{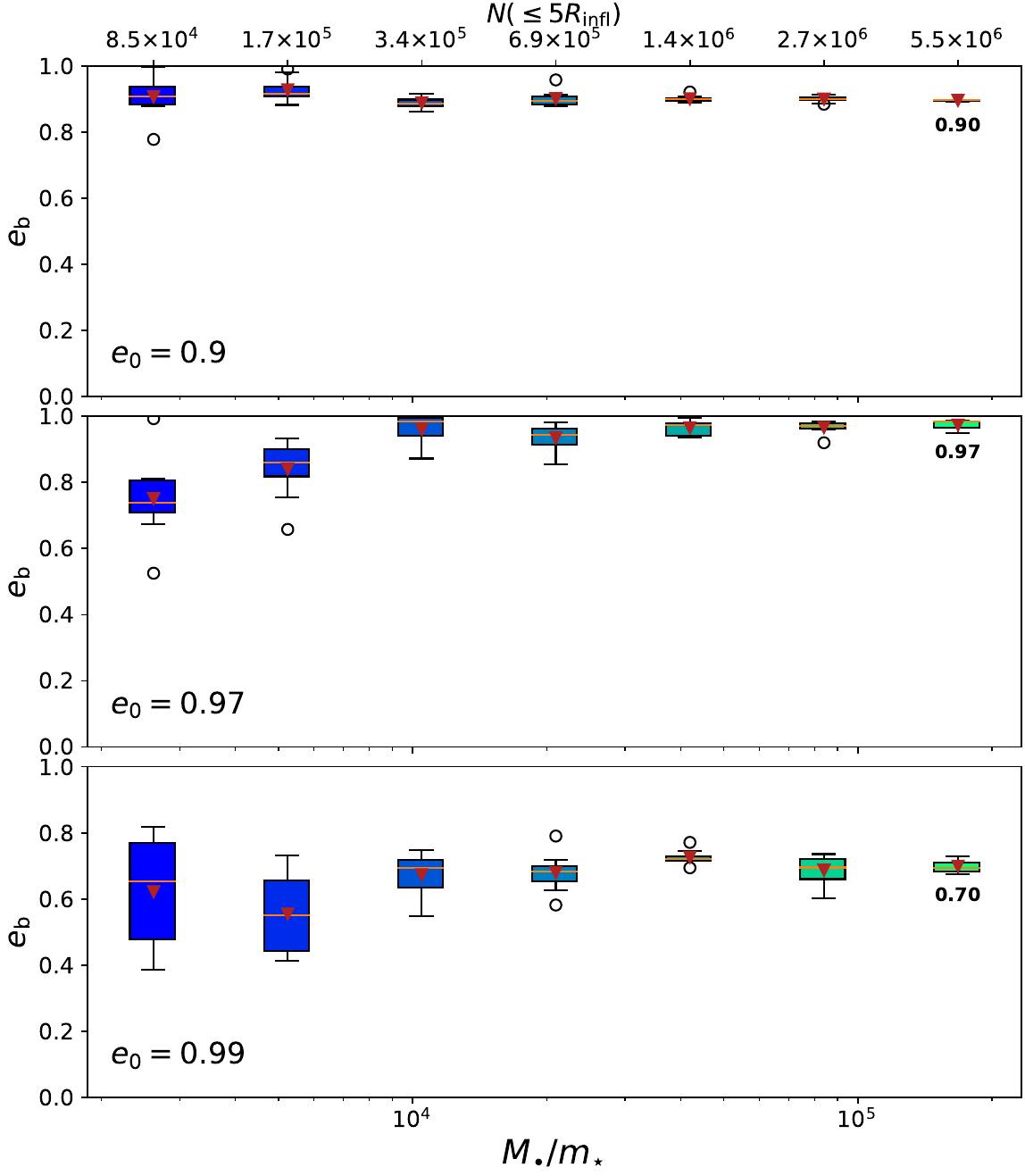}    
    \vspace*{-5mm}
    \caption{Box plot of the eccentricity at the time of binary formation as a function of resolution. The boxes extend from the first quartile to the third quartile of the data, with a horizontal line at the median and a triangle at the mean. The whiskers extend from the box to the farthest data point lying within 1.5 times the inter-quartile range from the box. The circles past the end of the whiskers indicate outliers. The colour gradient represents resolution. Convergence is reached at lower resolution for the lower eccentricity models than for the almost radial initial orbits. For the highest resolution, the average $\langle e_{\mathrm{b}}\rangle $ is given underneath the corresponding box.}
    \label{fig:boxplot}
\end{figure}

Fig.~\ref{fig:disp_ecc} does not show evidence of flattening at the largest resolution. We therefore explore at what resolution convergence is reached for models with different initial $e_0$. Fig.~\ref{fig:boxplot} shows a box-plot type distribution of eccentricity at the time of binary formation as a function of resolution, separately for the three different values of $e_0$. 
We define the point of convergence as the resolution where the average eccentricity $\langle e_{\mathrm{b}} \rangle$, marked by the triangle, changes by less than a given threshold tolerance (absolute difference) $h=0.02$ when doubling $N(< 5 R_{\rm infl})$. In other words, convergence is achieved when this difference first falls below the specified tolerance $h$. The resulting resolution requirements are listed in Table~\ref{tab:convergence}, in terms of both $N(< 5 R_{\rm infl})$ and total particle number $N$.  The latter is model-dependent and is provided solely for completeness. We consider the former a more reliable measure of resolution that can be computed and compared for any kind of single or multi-component model. The average $\langle e_{\mathrm{b}} \rangle$ of the largest resolution model is given under the corresponding box. This can be considered the limiting eccentricity of a MBHB for a given initial $e_0$. We note that the average in non-converged models does not match the converged limiting eccentricity, therefore performing multiple random realisations at low resolution and averaging the results for $e_{\mathrm{b}}$ is not equivalent to performing runs at a sufficiently large resolution that ensures convergence.

\begin{table}
\begin{tabular}{lcc}
\hline 
$e_0$ & $N(< 5 R_{\rm infl})$ & $N$ \\
\hline 
0.9 & $7\times10^5$ & 8M \\
0.97 & $7\times10^5$ & 8M \\
0.99 & $3\times10^6$ & 32M \\
\hline
\end{tabular}
\caption{Resolution requirements for convergence in the eccentricity at binary formation, for a tolerance threshold $h=0.02$. The tolerance is defined as the absolute difference in the mean eccentricity $\langle e_{\mathrm{b}} \rangle$ between consecutive resolution models differing by a factor of two in particle number $N$. Convergence is achieved when this difference first falls below the specified tolerance.}
\label{tab:convergence}
\end{table}

\subsection{Relation between binary eccentricity and the eccentricity of the galactic orbit}
\begin{figure}
	\includegraphics[width=\columnwidth]{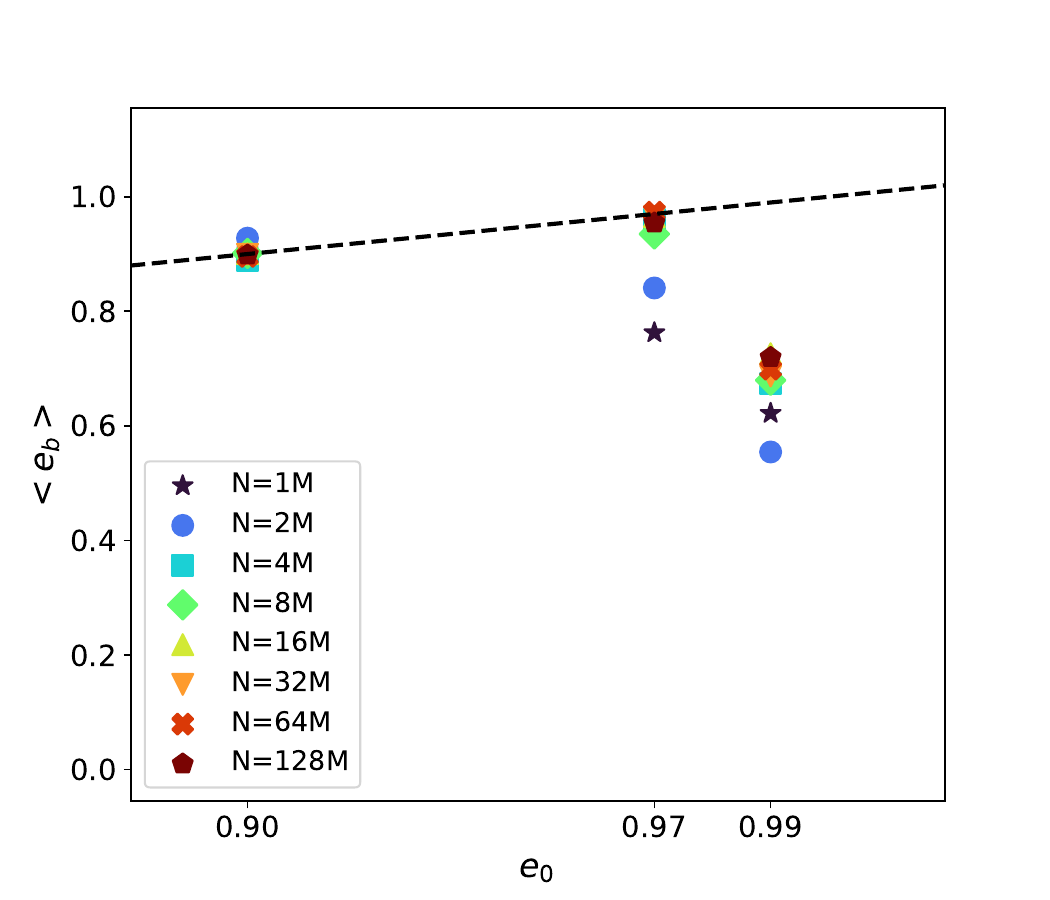}
    \vspace*{-5mm}
    \caption{Relation between the average eccentricity at binary formation and the initial eccentricity of the merger. The models with $e_0\leq0.97$ fit right on the relation of \citet{Gualandris2022}, though only at the larger resolutions for $e_0=0.97$, while the model with $e_0=0.99$ lies below the relation. There is a significant dependence on resolution, with larger $N$ models lying closer to the relation, but converging to a value of about $e_{\mathrm{b}}\sim0.7$. Additionally, the spread at a fixed resolution is larger for almost radial orbits.}
    \label{fig:relation_mean}
\end{figure}

We show the average eccentricity at binary formation as a function of $e_0$ and for different total $N$ in Fig.~\ref{fig:relation_mean}. Here the dashed line indicates the identity relation $e_{\mathrm{b}}=e_0$. We find that the models with $e_0\leq0.97$ converge to the identity relation of \citet{Gualandris2022},  while the models with $e_0=0.99$ converge to a value below the curve of about $e_{\mathrm{b}}=0.7$. Convergence is slower for more radial orbits, and this is already apparent at $e_0=0.97$.

\begin{figure*}
 \includegraphics[width=1.03\columnwidth]{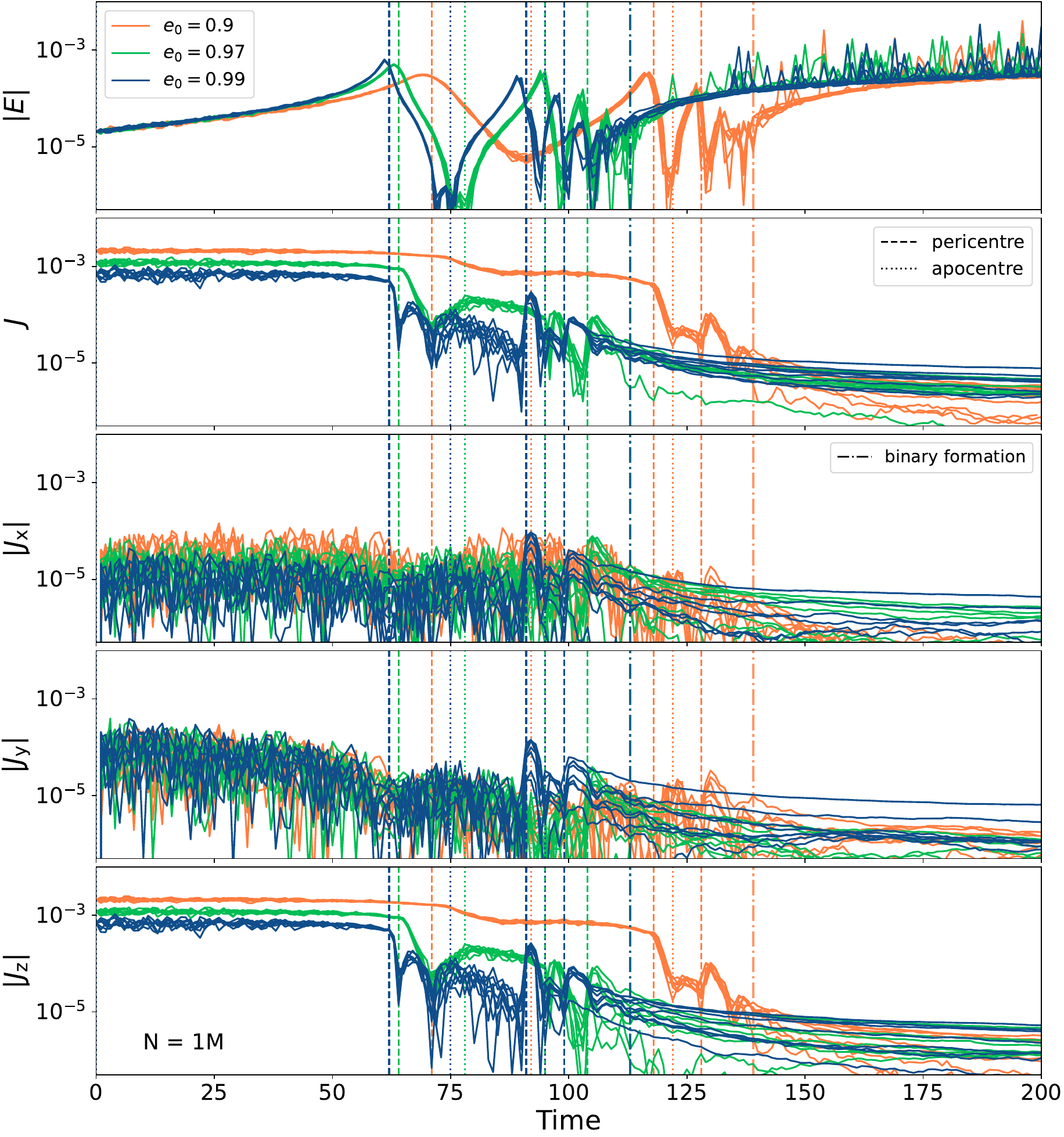}
 \hfill
 \includegraphics[width=1.03\columnwidth]{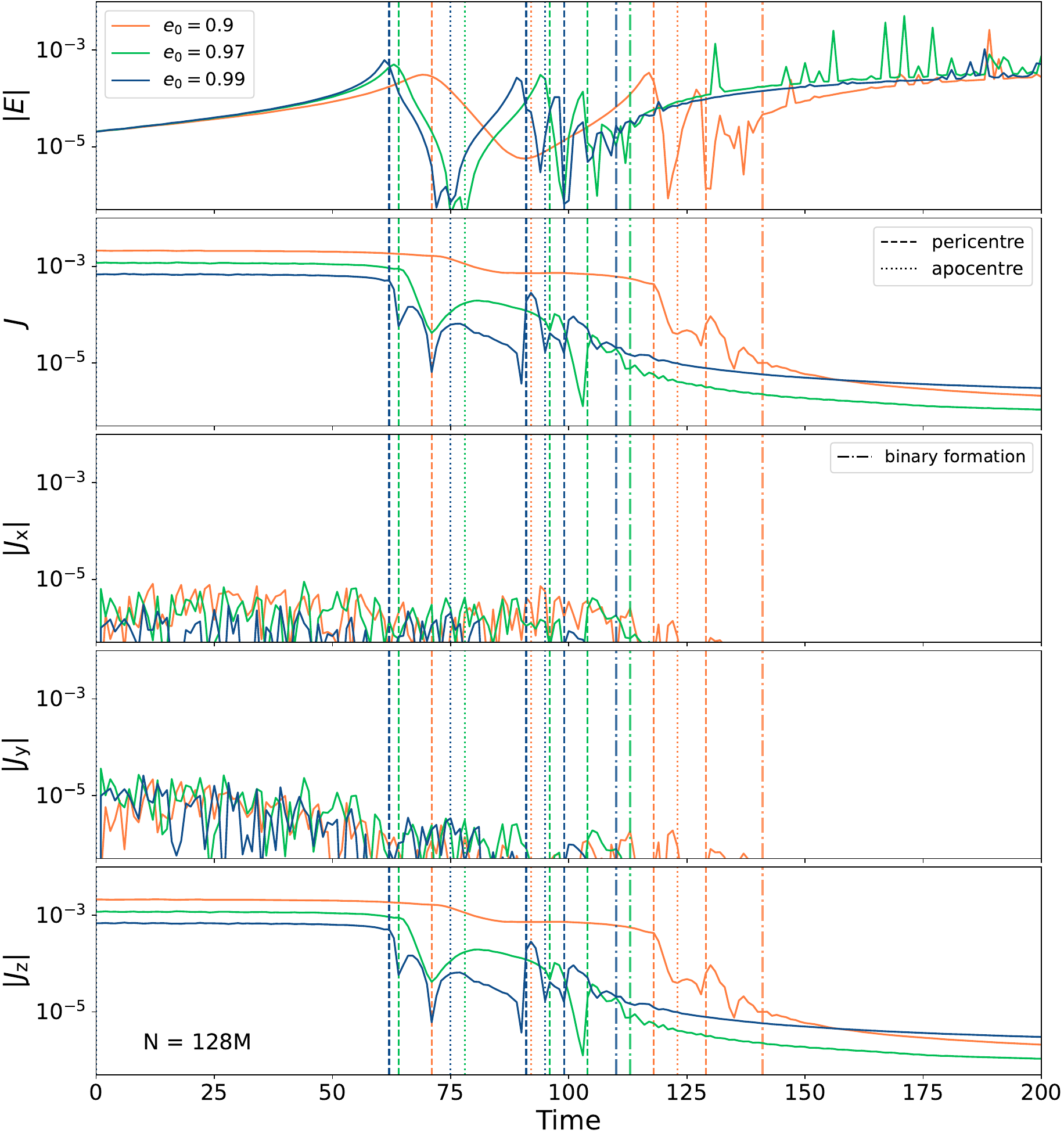}
    \vspace*{-2mm}
    \caption{Total energy $E$, angular momentum $J$ and its components of the MBH pairs as a function of time at low (left) and high (right) resolution, for the three different values of the initial orbital eccentricity $e_0$. The angular momentum decreases at each pericentre passage, and in particular the first, with a larger change for the more eccentric orbits. Numerical stochasticity arises after the first pericentre passage, visible in the spread of $J$ values at low resolution. This implies a spread of eccentricity by the time of binary formation. 
    }
    \label{fig:ej}
\end{figure*}
The break from the identity relation of \citet{Gualandris2022} for $e_0 > 0.97$ owes to angular momentum transfer in the early phases of the merger, and in particular at pericentre passages, and how such changes compare with the orbital angular momentum of the MBHs. More radial orbits are characterised by earlier and deeper pericentre passages in the potential of the galactic host. We show the time evolution of the energy and angular momentum of the MBHBs as a function of $e_0$ in Fig.~\ref{fig:ej}, both at low ($N=1\mathrm{M}$) and high ($N=128\mathrm{M}$) resolution. We assume a reference frame where the orbits of the initial galaxies lie in the $x-y$ plane, so that the initial angular momentum is aligned with the $z$-axis. 
The angular momentum $J$ is approximately constant until the first pericentre passage, which supports the finding in Fig.~\ref{fig:orb_elements} that the orbital eccentricities of the MBH pairs initially match $e_0$. It is then found to decrease sharply at pericentre passages, especially the first, with a larger change for the more eccentric orbits. Because the orbits with $e_0=0.99$ result in earlier and deeper pericentre passages (marked by the vertical dashed lines), larger changes in angular momentum have taken place by the time of binary formation (marked by the dashed-dotted lines). 

The angular momentum $J$ is approximately constant until the first pericentre passage, which supports the finding in Fig.~\ref{fig:orb_elements} that the orbital eccentricities of the MBH pairs initially match $e_0$. It is then found to decrease sharply at pericentre passages, especially the first, with a larger change for the more eccentric orbits. Because the orbits with $e_0=0.99$ result in earlier and deeper pericentre passages (marked by the vertical dashed lines), larger changes have taken place by the time of binary formation (marked by the dashed-dotted lines). This results in a larger spread of eccentricity values at binary formation.
We also note that numerical stochasticity arises after the first pericentre passage, as apparent in the low resolution simulations. This applies to all eccentricities, but is more pronounced for larger $e_0$. Changes in $J_z$, and therefore in $J$, lead to a spread in eccentricity at the time of binary formation, that decreases with increasing resolution.

\begin{figure*}
 \includegraphics[width=1.03\columnwidth]{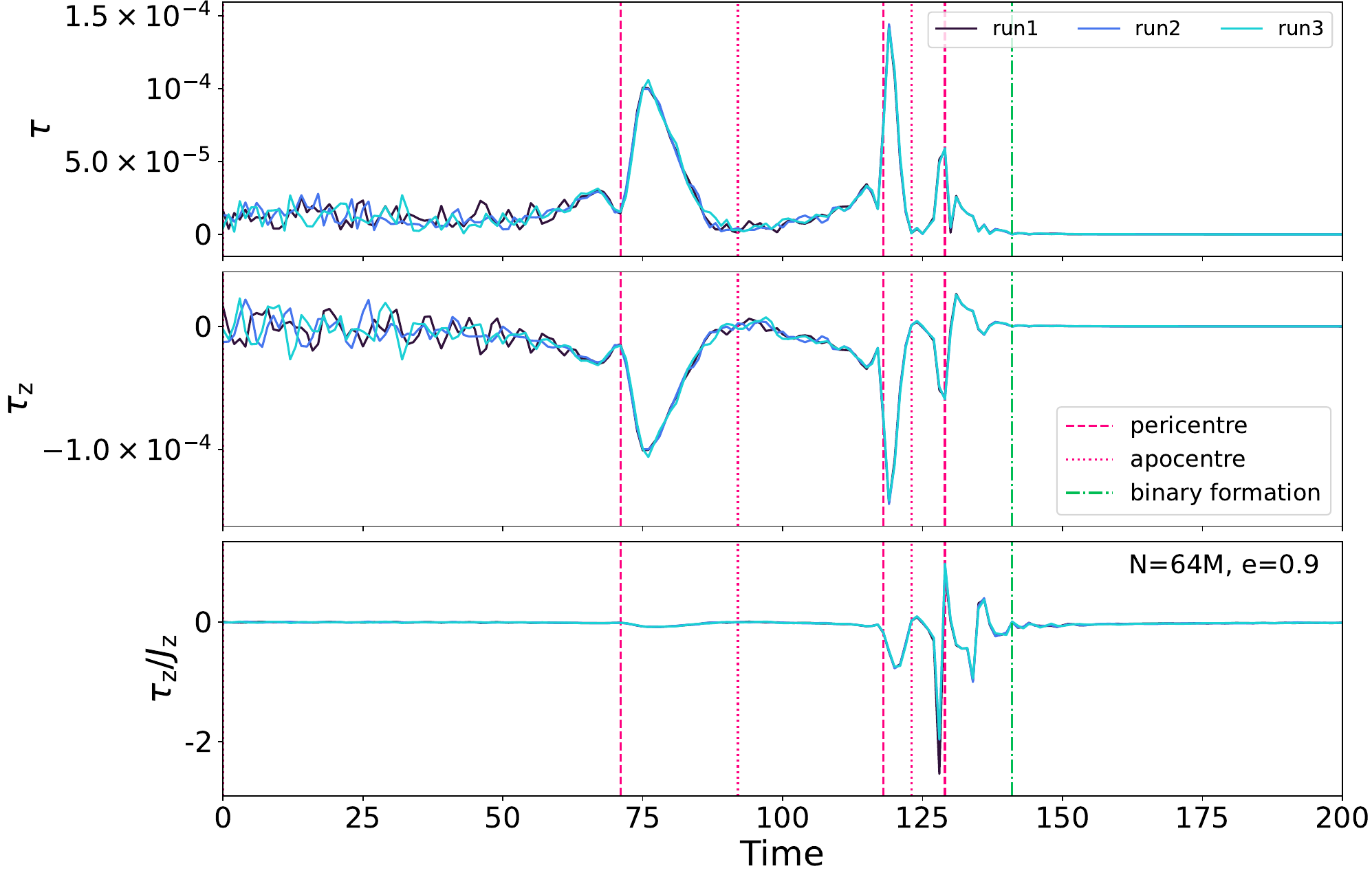}
 \hfill
 \includegraphics[width=1.03\columnwidth]{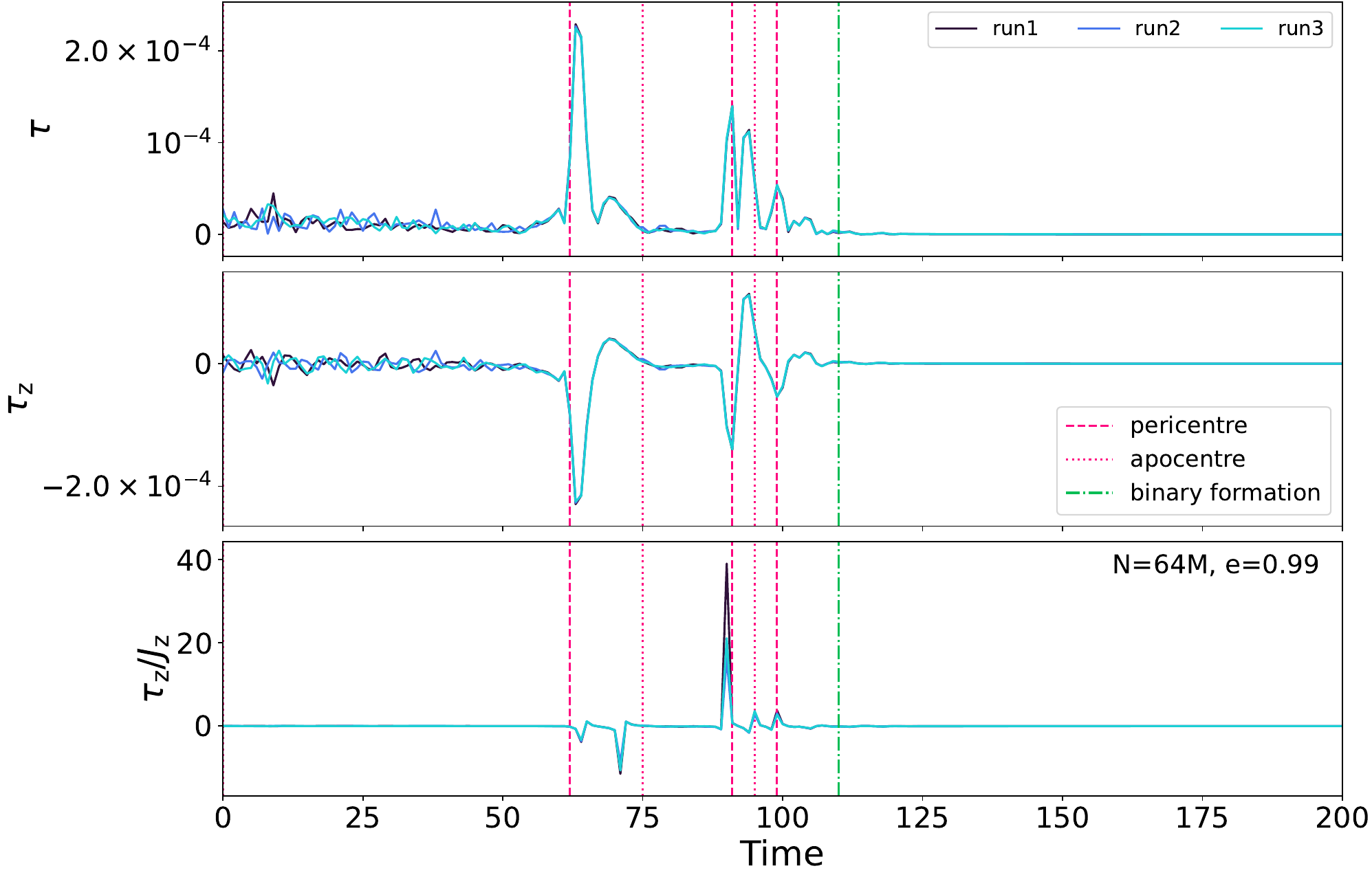}
    \vspace*{-2mm}
    \caption{Torque experienced by the MBH pair as a function of time in a high resolution ($N=64\mathrm{M}$) set of runs with initial orbital eccentricity $e_0=0.9$ (left panel) and $e_0=0.99$ (right panel). The bottom panel shows $\tau_z/J_z = d\ln J_z/dt$, which is directly linked to the variation in eccentricity. The MBHBs experience a torque at pericentre passages due to asphericities in the stellar distribution. The torque is larger at the first passage and for the more eccentric orbits, resulting in a larger change in angular momentum. }
    \label{fig:torque}
\end{figure*}
The angular momentum changes experienced by the binaries at pericentre are due to torques from the non-spherical stellar distribution. 
These are plotted in Fig.~\ref{fig:torque} for a high resolution ($N$=64M) suite, comparing the lowest $e_0=0.9$ and highest initial eccentricity $e_0=0.99$. Torques are generally small but exhibit clear peaks at pericentre passages. Peak torques are larger for the more eccentric orbits, and their effect is to drive changes in angular momentum and, as a result, in eccentricity. 
These changes lead to a temporary increase in eccentricity, which can be seen in all models in Fig.~\ref{fig:orb_elements}, followed by a reduction in eccentricity prior to binary formation.
The overall effect of the torques on the properties of the MBH pair depends on the initial total angular momentum of the orbit, i.e. on $e_0$. 
\begin{figure}
 \includegraphics[width=\columnwidth]{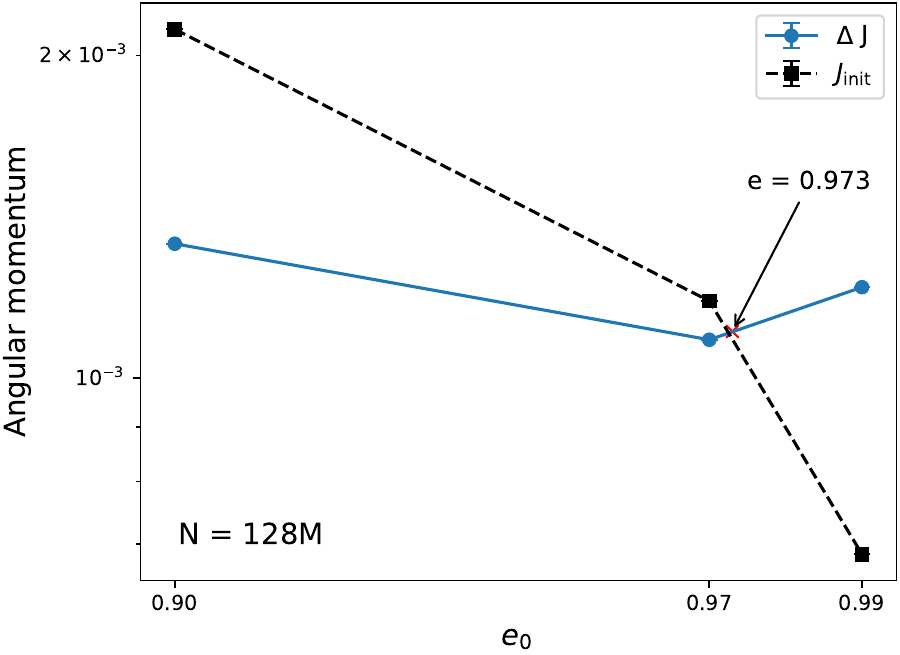}
    \vspace*{-5mm}
    \caption{Comparison between the total change in angular momentum $\Delta J$ produced by the torque from the non-spherical stellar distribution and the initial total angular momentum of the MBHB $J_{\rm init}$ as a function of the merger eccentricity $e_0$. While $\Delta J$ increases (albeit slowly) with $e_0$, $J_{\rm init}$ decreases for more radial orbits. There is a cross-over point around $e_0=0.97$ above which the torque results in a circularisation of the MBHB.}
    \label{fig:jdot}
\end{figure}
We compute the total change in angular momentum $\Delta J$ experienced by MBHBs in different models as a function of $e_0$, and compare it with the initial angular momentum $J_{\rm init}$ of the pair. These are shown in Fig.~\ref{fig:jdot} for the highest resolution simulations ($N$=128M). The change $\Delta J$ is lower than $J_{\rm init}$ for $e_0\leq 0.97$ but larger for $e_0=0.99$. This implies that there exists a cross-over eccentricity $e_{\mathrm{c}}$ above which the torques result in a circularisation of the binary. In this case, the eccentricity $e_{\mathrm{b}}$ of the binary at formation is lower than the initial $e_0$. This explains the turnover of the limiting distribution of $e_{\mathrm{b}}$ seen in Fig.~\ref{fig:relation_mean} above $e_0\sim 0.97$. The existence of a turnover critical eccentricity is a general result that can be expected for any model, regardless of the nature of the density profile. The precise value of $e_{\mathrm{c}}$, however, will depend on the choice of model. 

We perform a Markov chain Monte Carlo (MCMC) fit of the eccentricity at binary formation $e_{\mathrm{b}}$ as a function of the initial merger eccentricity $e_0$, adopting
uniform priors for the parameters and Gaussian errors for the likelihood function. We fit the following relation
\begin{equation}
e_{\mathrm{b}} =
\begin{cases}
  e_0, & \text{if } e_{\mathrm{b}} < e_{\mathrm{c}}, \\
  e_{\mathrm{c}} + (e_0 - e_{\mathrm{c}}) - H (e_0 - e_{\mathrm{c}})^2, & \text{else}
\end{cases}
\end{equation}
to the available data, using the average $e_{\mathrm{b}}$ over the random realisations for each model. We obtain the following best fit parameters: $e_{\mathrm{c}} = 0.966$ and $H = 478.88$, which agree very well with the location of the critical eccentricity found in Fig.~\ref{fig:jdot}.

\begin{figure*}
 \includegraphics[width=1.03\columnwidth]{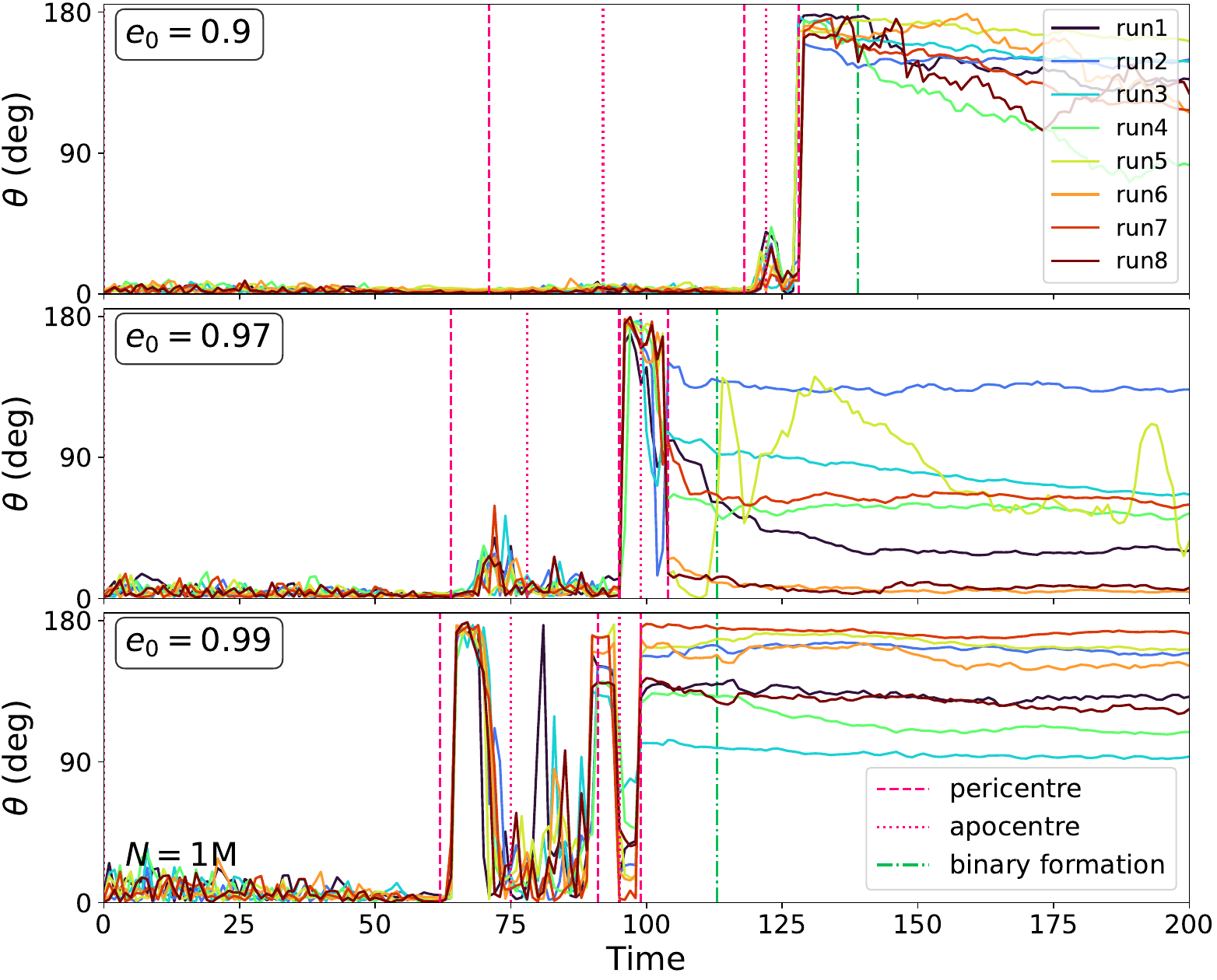}
 \hfill
 \includegraphics[width=1.03\columnwidth]{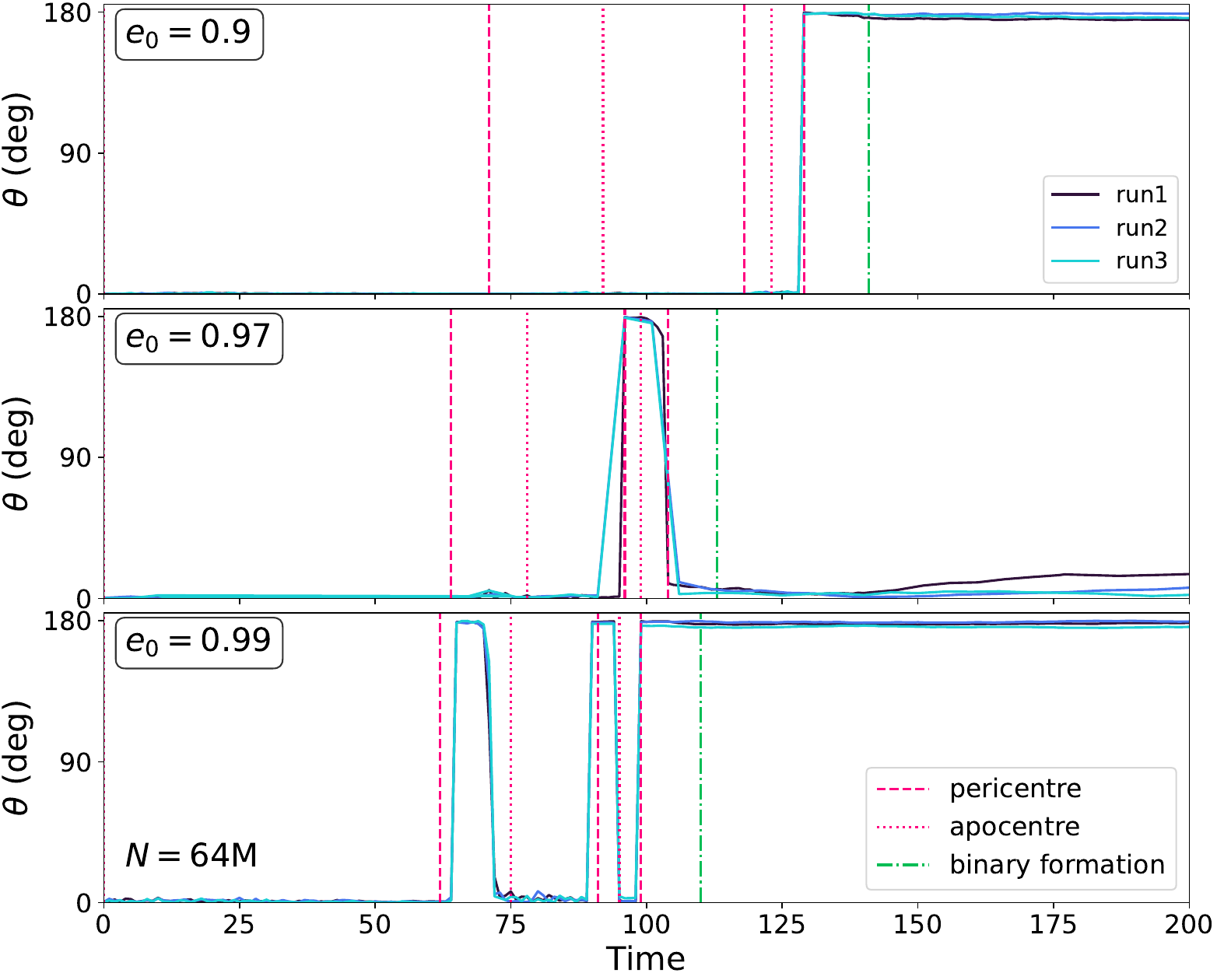}
    \vspace*{-2mm}
    \caption{Angle between the total angular momentum vector of the MBHB and that of the stars, as a function of time, for the three different values of the initial orbital eccentricities. (Left) Low resolution runs ($N=1\mathrm{M}$) (Right) High resolution runs ($N=64\mathrm{M}$). Flips of the orbital plane occur more often for more eccentric orbits, in correspondence with pericentre passages. Numerical stochasticity manifests itself also as a spread in the inclination angle of the MBHB after binary formation.}
    \label{fig:angle}
\end{figure*}

\subsection{Orbit flips}
Angular momentum changes in low $J_{\rm init}$ orbits also lead to spin flips, i.e. flips in the direction of rotation with respect to the stellar system.  Flips have been previously reported in the literature in different contexts.  Deep, nearly head-on encounters of massive halos can lead to a reversal of the direction of orbital motion of the central galaxies after their first pericentre passage \citep{Barnes2016}. Major mergers of massive early-type galaxies with central MBHs can lead to the reversal of the MBH orbits due to gravitational torques following pericentre passages \citep{Rantala2019}. In rotating galaxies, BHBs on retrograde orbits flip to corotating orbits to align with the angular momentum of the galaxy after formation \citep{Khan2020}. 
 
Fig.~\ref{fig:angle} shows the angle between the angular momentum vector of the MBHB and the total angular momentum vector of the stars for the three different values of $e_0$, at both low and high resolution. We find that flips occur in all cases, but while only one flip occurs for $e_0$, five flips occur for $e_0=0.99$. Low angular momentum orbits are more likely to flip their rotation with respect to the background when subject to torques from the non-spherical merger remnant.  Flips happen after pericentre passages, when torques peak. However, only one passage is required to flip the $e_0=0.99$ orbit due to its low initial $J_{\rm init}$, followed by four more flips, and the binary remains flipped compared to the initial configuration after binding. For $e_0=0.9$, on the other hand, a flip occurs only after the third pericentre passage, and the binary also remains flipped after binding. In the intermediate $e_0=0.97$ case, the binary flips after the second pericentre passage, but returns to its original orientation. The effects of numerical stochasticity are also seen in the angle $\theta$ between the total angular momentum vector of the MBHB and that of the stars after binary formation, with a significant spread at low resolution. Our results are consistent with those of \citet{Gualandris2022}, who show that flips occur in models with large orbital eccentricity and a shallow central density profile. These models are characterized by a mildly oblate or triaxial shape in the innermost regions \citep{Bortolas2018}, which gives rise to torques acting on the MBHB. Importantly, spin flips correlate with the largest variations in $J_z$ and therefore in eccentricity.

\subsection{Merger timescales}
\label{sec:times}
We explore the effects of numerical and physical stochasticity on the estimated merger timescales of MBHBs. First, we extend the modelling of the binary evolution obtained with the $N$-body simulations past the phases of formation and hardening by means of a semi-analytical model similar to that described in \citet{attard2024}. The model solves the system of coupled differential equations for the binary semi-major axis $a$ and eccentricity $e$:
\begin{align}
    \dv{a}{t} &= \eval{\dv{a}{t}}_{*}  + \eval{\dv{a}{t}}_{\mathrm{GW}}, \label{eq:dadt} \\
    \dv{e}{t} &= \eval{\dv{e}{t}}_{*}  + \eval{\dv{e}{t}}_{\mathrm{GW}}, \label{eq:dedt}
\end{align}
where the first term represents hardening from three-body encounters with stars and the second term represents losses due to GW emission. 
The stellar hardening is included in the $N$-body model, and we compute the associated time-dependent hardening rate 
\begin{align}
	s(t) &= \dv{t} \qty(\frac1a) = - \frac1{a^2} \eval{\dv{a}{t}}_*
\end{align}
and eccentricity growth rate
\begin{align}
	K(t) &= \dv{e}{\ln(1/a)}  = \frac1{sa} \eval{\dv{e}{t}}_*
\end{align}
directly from the $N$-body data (rather than estimate them based on scattering experiments).

The contribution to orbital decay from GW emission of two point mass black holes $M_1$ and $M_2$ is modelled with the formalism presented in \citet{peters1964}:
\begin{align}
    \eval{\dv{a}{t}}_{\mathrm{GW}}
    &= -\frac{64}{5}\frac{G^{3}M_{1}M_{2}(M_1+M_2)}{c^{5}a^{3}(1-e^{2})^{7/2}}\left( 1+\frac{73}{24}e^{2}+\frac{37}{96}e^{4}\right),
	\label{eq:peters_a}
    \\
    \eval{\dv{e}{t}}_{\mathrm{GW}}
    &= -\frac{304}{15}e\frac{G^{3}M_{1}M_{2}(M_1+M_2)}{c^{5}a^{4}(1-e^{2})^{5/2}}\left( 1+\frac{121}{304}e^{2} \right).
	\label{eq:peters_e}
\end{align}
These equations show the strong scaling of the orbital elements with the binary eccentricity.

We start the semi-analytical model at a time chosen during the early stages of hardening in order to be able to validate the predicted evolution against the $N$-body simulation before relying on the inferred hardening and eccentricity growth rates for forecasting to late times. We measure the merger time as the total time elapsed between the simulated galaxy merger and the coalescence of the MBHs.

We compute the time-dependent hardening rate $s(t)$ and eccentricity growth rate $K(t)$ through direct numerical differentiation of the orbital elements in small time intervals. 
The orbital parameters, particularly the eccentricity, may be poorly defined near pericentre passages due to rapid variations in separation and velocity. To obtain robust estimates of the secular evolution, we therefore compute the rates from cleaned orbital data, i.e. after removing pericentre passages. 

We fit an exponential relation to $s$ and a linear relation to $K$, as these provide the best match to the $N$-body data, and we then use these for the extrapolation to late times. 

In order to compute GW losses and total merger times, we need to assign physical scales to our scale-free systems. We adopt a range of physical MBH masses typical of PTA sources, $M_{\rm bh} = (10^7-10^{10})\msun$, and set the physical length scale to match the radius of influence of the MBH. This is computed assuming a velocity dispersion from the $M_{\rm bh}-\sigma$ relation of \citet{msigma2009}. The corresponding time and velocity units are given by $[T] = \sqrt{[L]^3/(G[M])}$ and $[V]=\sqrt{G[M]/[L]}$.

\begin{table}
\begin{tabular}{lcccc}
\hline 
$M_{\rm bh}/\msun$ & $e_0$  & $T_{\rm m}$ (yr) LR & $T_{\rm m}$ (yr) HR &
$|T_{\rm m}\,{\rm HR} - T_{\rm m}\,{\rm LR}|$ (yr)
\\
\hline 
$10^7$ & 0.9 & $8.4\times10^6$ & $2.0\times10^7$ & $1.1\times10^7$ \\
$10^8$ & 0.9 & $1.5\times10^7$ & $1.8\times10^7$ & $3.4\times10^6$\\
$10^9$ & 0.9 & $2.7\times10^7$ & $3.0\times10^7$ & $2.8\times10^6$ \\
$10^{10}$ & 0.9 & $5.1\times10^7$ & $5.4\times10^7$ & $3.3\times10^6$\\
\hline
$10^7$ & 0.99 & $4.8\times10^8$ & $1.1\times10^8$ & $3.7\times10^8$ \\
$10^8$ & 0.99 & $5.9\times10^8$ & $4.4\times10^7$ & $5.4\times10^8$\\
$10^9$ & 0.99 & $1.2\times10^8$ & $4.8\times10^7$ & $7.6\times10^7$ \\
$10^{10}$ & 0.99 & $8.5\times10^7$ & $6.9\times10^7$ & $1.5\times10^7$\\
\hline
\end{tabular}
\caption{Merger timescales $T_{\rm m}$ for the MBHBs, averaged over the available random realisations, at low resolution (LR, $N$=1M) and high resolution (HR, $N=64\mathrm{M}$), for $e_0=0.9, 0.99$, and the absolute difference between the LR and HR merger times. Merger times depend significantly on the adopted physical scaling, and exhibit larger discrepancies for more eccentric orbits.}
\label{tab:merger_times}
\end{table}
In Table \ref{tab:merger_times} we list the derived merger timescales $T_{\rm m}$ for models with $e_0=0.9$ and $e_0=0.99$ for two distinct resolutions: $N=1\mathrm{M}$ (LR) and $N=64\mathrm{M}$ (HR). The estimates are averages over the number of available random realisations at each resolution. The last column in the table gives the absolute difference in $T_{\rm m}$ between the low and high resolution values. Merger times depend significantly on the adopted physical scaling, with the more massive black holes generally characterised by longer times to coalescence, but with a strong dependence on the eccentricity at binary formation and hardening and eccentricity growth rates. Furthermore, the effect of resolution on merger times is significant, with a relative error ranging from $6\%$ to $57\%$ for $e_0=0.9$ and from $22\%$ to $350\%$ for $e_0=0.99$, depending on MBH mass. In comparison, the uncertainties due to numerical differentiation in the computation of $s$ and $K$ result in relative differences in $T_{\rm m}$ ranging from $10\%$ for the HR models to a factor 2 for the LR models.

\begin{figure}
 \includegraphics[width=\columnwidth]{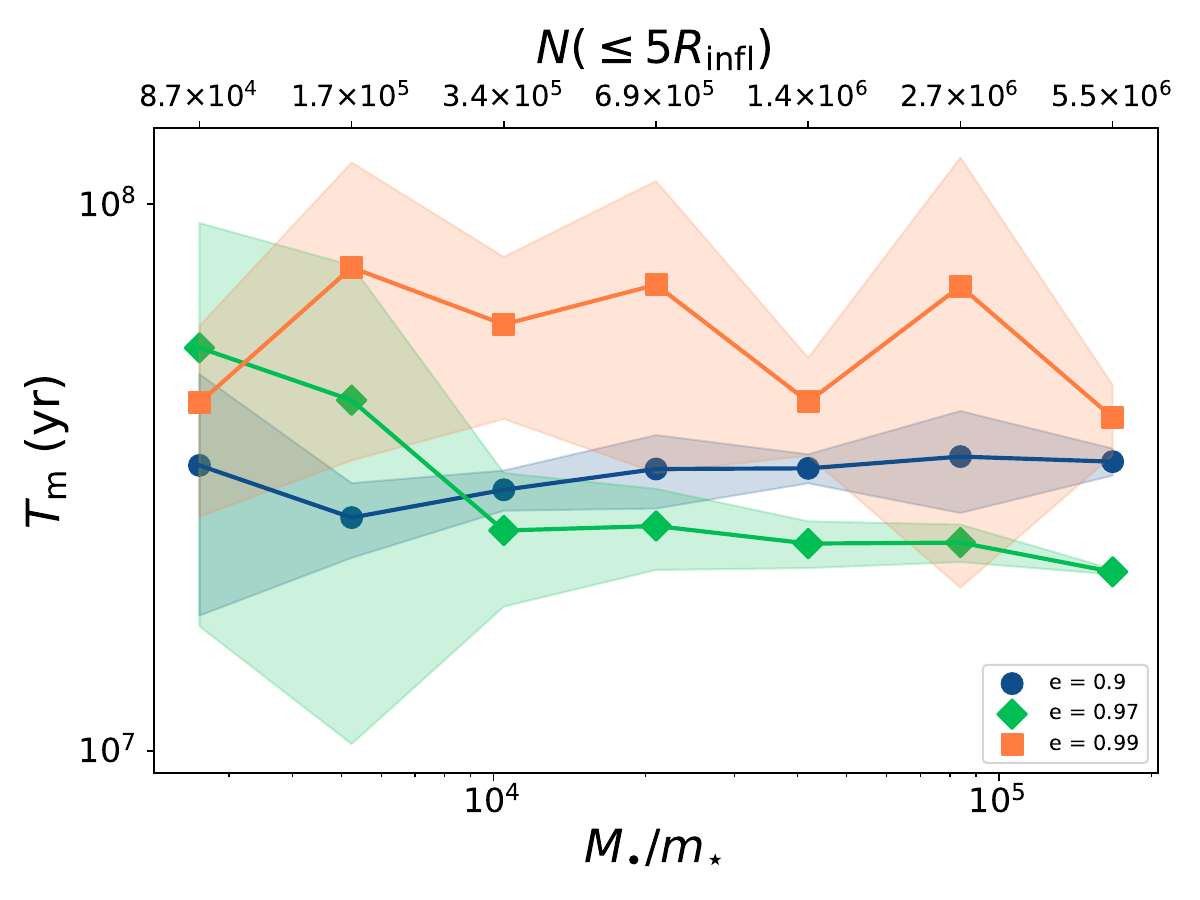}
    \vspace*{-5mm}
    \caption{Merger timescale for the MBHBs (from galaxy merger to GW coalescence of the MBHs) as a function of resolution, assuming $M_{\bullet} = 10^9\msun$. The points represent the average timescale over the available random realisations for each model and the shaded areas represent the $1\sigma$ confidence band.}
    \label{fig:tmerg_ave}
\end{figure}
We show the average merger timescale $T_{\rm m}$ as a function of resolution for the $M_{\bullet} = 10^9\msun$ scaling in Fig.~\ref{fig:tmerg_ave}, with $1\sigma$ confidence bands given by the dispersion over the available random realisations for each model. We note that models with $e_0=0.97$ achieve, on average, the shortest merger times. This is due to the fact that these models form the most eccentric binaries. As shown in Fig.~\ref{fig:relation_mean}, mergers with initial eccentricity larger than the critical $e_{\mathrm{c}} \sim 0.97$ form binaries with lower eccentricity than expected based on the identity relation. On the other hand, given their lower $e_{\mathrm{b}}\sim 0.7$,  binaries formed in mergers with $e_0=0.99$ have an average merger time which is longer than binaries formed in mergers with $e_0=0.9$. Numerical stochasticity has a significant effect on the merger timescales that reduces with increasing resolution. The required resolution is larger for more eccentric galactic encounters. 

\subsection{Resolution requirements}
We present a fitting formula for the resolution requirement of an $N$-body simulation of MBHB formation and evolution as a function of the initial eccentricity of the merger and the required accuracy in the eccentricity of the binary formation. The number of particles $N(\leq 5 R_{\rm infl})$ within 5 influence radii of the primary MBH required to achieve a desired dispersion $\sigma_e$ in $e_{\mathrm{b}}$ is: 
\begin{equation}
    N(\leq 5 R_{\rm infl}) = \left( \frac{\sigma_e}{p}  {e_0}^{b}\right)^{1/c}
\end{equation}
where $b, c, p$ are numerical factors to be determined from fitting the data in Fig.~\ref{fig:disp_ecc}. We perform a MCMC fit of the dispersion in eccentricity as a function of resolution adopting uniform priors for the parameters and Gaussian errors for the likelihood function. We obtain the following best fit parameters with $68\%$ uncertainties: $b = 12.79^{+1.05}_{-1.16}$,  $c = -0.63^{+0.03}_{-0.04}$ and $p = 2.79^{+3.3}_{-1.28}$.  This result supports the expectation for a Poisson-like scaling with resolution, or a little steeper.
\begin{figure}
 \includegraphics[width=\columnwidth]{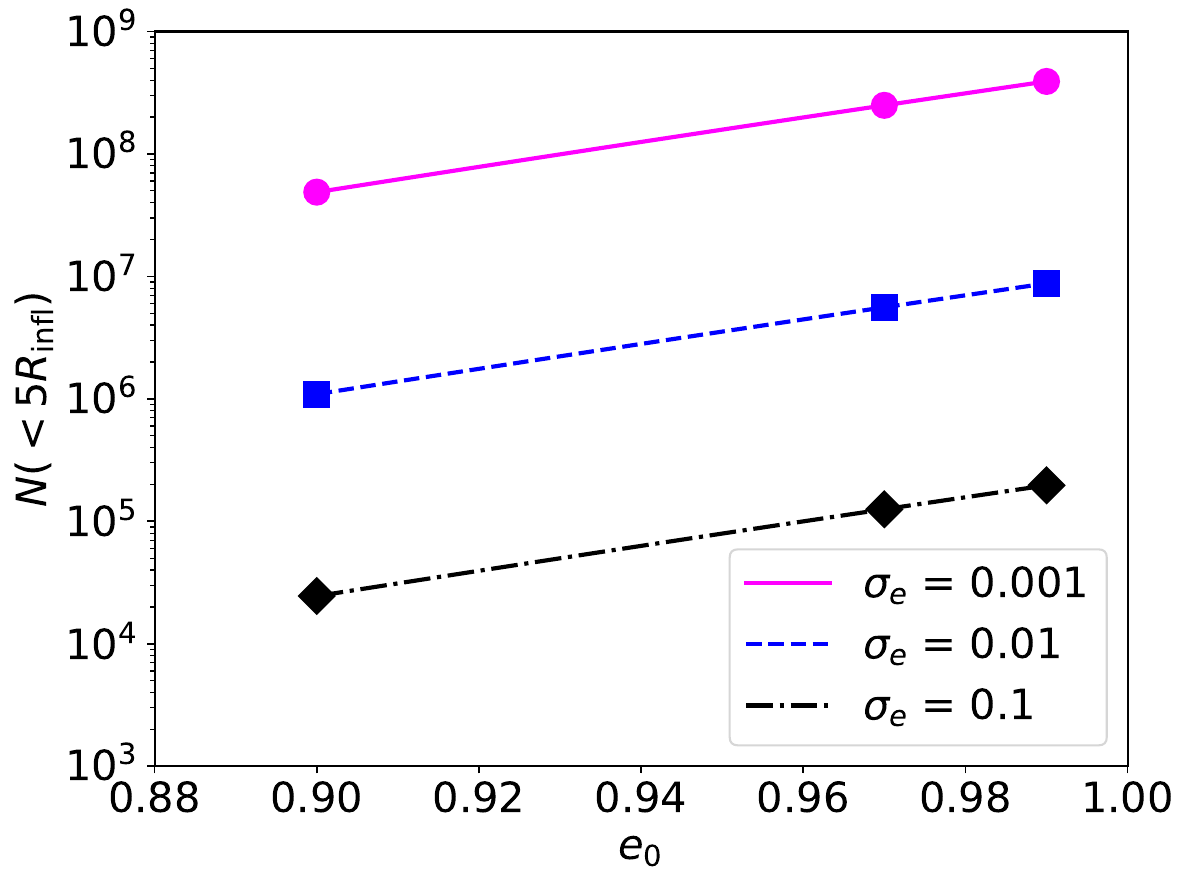}
    \vspace*{-5mm}
    \caption{Numerical resolution of an $N$-body simulation of MBH formation and evolution required to achieve a desired dispersion in eccentricity at binary formation as a function of the initial eccentricity of the merger, in the range $0.9-0.99$.}
    \label{fig:resolution}
\end{figure}
Fig.~\ref{fig:resolution} shows the requirements in the tested eccentricity range $0.9-0.99$ for three different values of $\sigma_e$. We caution that this fitting formula is only valid in the tested range of eccentricities $0.9-0.99$. We expect the relation to flatten for less eccentric binaries but additional simulations with low initial $e_0$ would be required to extend the relation into the low eccentricity regime.

\section{Discussion}
The simulation and analysis presented in this work focus on the high eccentricity range of galactic mergers. This is justified by the recent exploration of cosmological simulations by \citet{Fastidio2024} showing that the majority of mergers occur on nearly radial orbits, with a mean eccentricity of 0.88, a median eccentricity of 0.97, and with approximately 60\% of them having eccentricity $0.95 \leq e_0 < 1$.  Modelling of these galactic collisions in the context of MBH formation and evolution, e.g. for predictions of GW rates and the PTA signal, requires significantly higher resolution than is generally adopted, and which increases the more eccentric the orbit is. The resolution, expressed in particle number within a few times the influence radius of the primary MBH, while challenging from a computational point of view, is still very small compared to stellar populations in real galaxies. We therefore do not expect stochastic effects in real galaxies, unless massive perturbers significantly affect MBH formation and evolution. \citet{Rawlings2023} suggest that substructures in the galaxies may act as perturbers on the MBHBs, effectively randomising the eccentricity at binary formation. This is an intriguing possibility that requires further investigation.

We caution that our results and the fitting formulae derived for the eccentricity at binary formation and the resolution requirements apply directly only to the models presented here. These models have been chosen to represent typical hosts to PTA sources \citep{IzquierdoVillalba2023} but are idealised and sample only a narrow subset of possible merger configurations. Among massive ellipticals, the eccentricity at binary formation may depend on the mass ratio between the galaxies and between the MBHs and on the presence of a significant dark matter halo. For instance, \citet{Merritt2006} and \citet{Sesana2006} report a higher hardening rate and eccentricity growth rate for smaller mass ratios. Furthermore, the assumption of spherical galaxies may result in longer timescale than moderately triaxial systems.
Gas processes will significantly affect binary formation and evolution, therefore the modelling of gas-rich galaxies, particularly relevant to LISA sources, would require simulations including gas physics. 

The abundance of eccentric binaries has important implications for the astrophysical interpretation of the GWB signal from PTAs. The potential flattening in the GWB spectrum at low frequencies can be explained by environmental effects, including a population of eccentric binaries. In general, eccentric binaries at the onset of the GW inspiral phase originate from eccentric binaries at formation \citep[e.g.][]{Gualandris2022}, which in turn result from eccentric galactic mergers. However, the most eccentric binaries at formation come from mergers with $e_0 = e_{\mathrm{c}} \sim 0.97$, and not from nearly radial mergers. The latter tend to produce moderately eccentric binaries. The eccentricity at binary formation is the key factor determining the total merger time of a MBHB, with the shortest times to coalescence originating from mergers with orbital eccentricity near the critical value.

\section{Conclusions}
We investigated the origin and implications of stochasticity in $N$-body simulations of galactic mergers leading to the formation and hardening of massive black hole binaries which are potential sources of GWs for PTAs. Stochasticity due to numerical resolution shows a Poisson $1/\sqrt{N}$ scaling with particle number for any initial eccentricity of the galactic merger. However, the numerical dispersion in eccentricity of the binaries at formation increases for more radial mergers and leads to larger resolution requirements. We provide a fitting formula for the resolution requirement of an $N$-body simulation of MBHB formation and evolution as a function of the initial eccentricity of the merger, for a given desired accuracy in 
the binary eccentricity.  The eccentricity of binaries at formation follows the approximately identity relation presented in \citet{Gualandris2022} up to a critical orbital eccentricity ($e_{\mathrm{c}} \sim 0.97$ for the simulations presented here). Binaries receive an approximately constant torque at first pericentre which causes them to circularise when the initial orbital eccentricity exceeds the critical value $e_{\mathrm{c}}$, leading to a turnover in the relation.
We argue that this behaviour should be universal and largely independent of the initial conditions of the host galaxies. We provide a fitting formula
for $e_{\mathrm{b}}$ given an initial galactic orbit that can be used in semi-analytical models to determine the merger timescales of MBHBs as well as the amplitude and
slope of the GWB.

\section*{Acknowledgements}
AG acknowledges support from grant ST/Y002385/1.
The simulations were run on the Eureka2 HPC cluster at the University of Surrey.

\section*{Data Availability}
The authors will share the data underlying this article upon reasonable request.



\bibliographystyle{mnras}
\bibliography{biblio} 

\begin{thebibliography}{}
\makeatletter
\relax
\def\mn@urlcharsother{\let\do\@makeother \do\$\do\&\do\#\do\^\do\_\do\%\do\~}
\def\mn@doi{\begingroup\mn@urlcharsother \@ifnextchar [ {\mn@doi@}
  {\mn@doi@[]}}
\def\mn@doi@[#1]#2{\def\@tempa{#1}\ifx\@tempa\@empty \href
  {http://dx.doi.org/#2} {doi:#2}\else \href {http://dx.doi.org/#2} {#1}\fi
  \endgroup}
\def\mn@eprint#1#2{\mn@eprint@#1:#2::\@nil}
\def\mn@eprint@arXiv#1{\href {http://arxiv.org/abs/#1} {{\tt arXiv:#1}}}
\def\mn@eprint@dblp#1{\href {http://dblp.uni-trier.de/rec/bibtex/#1.xml}
  {dblp:#1}}
\def\mn@eprint@#1:#2:#3:#4\@nil{\def\@tempa {#1}\def\@tempb {#2}\def\@tempc
  {#3}\ifx \@tempc \@empty \let \@tempc \@tempb \let \@tempb \@tempa \fi \ifx
  \@tempb \@empty \def\@tempb {arXiv}\fi \@ifundefined
  {mn@eprint@\@tempb}{\@tempb:\@tempc}{\expandafter \expandafter \csname
  mn@eprint@\@tempb\endcsname \expandafter{\@tempc}}}

\bibitem[\protect\citeauthoryear{{Agazie} et~al.,}{{Agazie}
  et~al.}{2023}]{NANOGRAV2023}
{Agazie} G.,  et~al., 2023, \mn@doi [\apjl] {10.3847/2041-8213/acdac6}, \href
  {https://ui.adsabs.harvard.edu/abs/2023ApJ...951L...8A} {951, L8}

\bibitem[\protect\citeauthoryear{{Amaro-Seoane} et~al.,}{{Amaro-Seoane}
  et~al.}{2017}]{LISA2017}
{Amaro-Seoane} P.,  et~al., 2017, \mn@doi [arXiv e-prints]
  {10.48550/arXiv.1702.00786}, \href
  {https://ui.adsabs.harvard.edu/abs/2017arXiv170200786A} {p. arXiv:1702.00786}

\bibitem[\protect\citeauthoryear{{Attard}, {Gualandris}, {Read}  \&
  {Dehnen}}{{Attard} et~al.}{2024}]{attard2024}
{Attard} K.,  {Gualandris} A.,  {Read} J.~I.,   {Dehnen} W.,  2024, \mn@doi
  [\mnras] {10.1093/mnras/stae524}, \href
  {https://ui.adsabs.harvard.edu/abs/2024MNRAS.529.2150A} {529, 2150}

\bibitem[\protect\citeauthoryear{{Barnes}}{{Barnes}}{2016}]{Barnes2016}
{Barnes} J.~E.,  2016, \mn@doi [\mnras] {10.1093/mnras/stv2381}, \href
  {https://ui.adsabs.harvard.edu/abs/2016MNRAS.455.1957B} {455, 1957}

\bibitem[\protect\citeauthoryear{{Begelman}, {Blandford}  \& {Rees}}{{Begelman}
  et~al.}{1980}]{Begelman1980}
{Begelman} M.~C.,  {Blandford} R.~D.,   {Rees} M.~J.,  1980, \mn@doi [\nat]
  {10.1038/287307a0}, \href {http://adsabs.harvard.edu/abs/1980Natur.287..307B}
  {287, 307}

\bibitem[\protect\citeauthoryear{{Benson}}{{Benson}}{2005}]{Benson2005}
{Benson} A.~J.,  2005, \mn@doi [\mnras] {10.1111/j.1365-2966.2005.08788.x},
  \href {https://ui.adsabs.harvard.edu/abs/2005MNRAS.358..551B} {358, 551}

\bibitem[\protect\citeauthoryear{{Bortolas}, {Gualandris}, {Dotti}  \&
  {Read}}{{Bortolas} et~al.}{2018}]{Bortolas2018}
{Bortolas} E.,  {Gualandris} A.,  {Dotti} M.,   {Read} J.~I.,  2018, \mn@doi
  [\mnras] {10.1093/mnras/sty775}, \href
  {https://ui.adsabs.harvard.edu/abs/2018MNRAS.477.2310B} {477, 2310}

\bibitem[\protect\citeauthoryear{{Dehnen}}{{Dehnen}}{1993}]{Dehnen1993}
{Dehnen} W.,  1993, \mn@doi [\mnras] {10.1093/mnras/265.1.250}, \href
  {https://ui.adsabs.harvard.edu/abs/1993MNRAS.265..250D} {265, 250}

\bibitem[\protect\citeauthoryear{{Dehnen}}{{Dehnen}}{2014}]{Dehnen2014}
{Dehnen} W.,  2014, \mn@doi [Computational Astrophysics and Cosmology]
  {10.1186/s40668-014-0001-7}, \href
  {https://ui.adsabs.harvard.edu/abs/2014ComAC...1....1D} {1, 1}

\bibitem[\protect\citeauthoryear{{EPTA Collaboration} et~al.,}{{EPTA
  Collaboration} et~al.}{2023}]{EPTA2023search}
{EPTA Collaboration} et~al., 2023, \mn@doi [\aap]
  {10.1051/0004-6361/202346844}, \href
  {https://ui.adsabs.harvard.edu/abs/2023A&A...678A..50E} {678, A50}

\bibitem[\protect\citeauthoryear{{EPTA Collaboration} et~al.,}{{EPTA
  Collaboration} et~al.}{2024}]{EPTA2023interp}
{EPTA Collaboration} et~al., 2024, \mn@doi [\aap]
  {10.1051/0004-6361/202347433}, \href
  {https://ui.adsabs.harvard.edu/abs/2024A&A...685A..94E} {685, A94}

\bibitem[\protect\citeauthoryear{{Fastidio}, {Gualandris}, {Sesana}, {Bortolas}
   \& {Dehnen}}{{Fastidio} et~al.}{2024}]{Fastidio2024}
{Fastidio} F.,  {Gualandris} A.,  {Sesana} A.,  {Bortolas} E.,   {Dehnen} W.,
  2024, \mn@doi [\mnras] {10.1093/mnras/stae1411}, \href
  {https://ui.adsabs.harvard.edu/abs/2024MNRAS.532..295F} {532, 295}

\bibitem[\protect\citeauthoryear{{Fastidio}, {Bortolas}, {Gualandris},
  {Sesana}, {Read}  \& {Dehnen}}{{Fastidio} et~al.}{2025}]{Fastidio2025}
{Fastidio} F.,  {Bortolas} E.,  {Gualandris} A.,  {Sesana} A.,  {Read} J.,
  {Dehnen} W.,  2025, \mn@doi [arXiv e-prints] {10.48550/arXiv.2506.16539},
  \href {https://ui.adsabs.harvard.edu/abs/2025arXiv250616539F} {p.
  arXiv:2506.16539}

\bibitem[\protect\citeauthoryear{{Ferrarese} \& {Merritt}}{{Ferrarese} \&
  {Merritt}}{2000}]{FerrareseMerritt2000}
{Ferrarese} L.,  {Merritt} D.,  2000, \mn@doi [\apjl] {10.1086/312838}, \href
  {https://ui.adsabs.harvard.edu/abs/2000ApJ...539L...9F} {539, L9}

\bibitem[\protect\citeauthoryear{{Gualandris} \& {Merritt}}{{Gualandris} \&
  {Merritt}}{2012}]{GualandrisMerritt2012}
{Gualandris} A.,  {Merritt} D.,  2012, \mn@doi [\apj]
  {10.1088/0004-637X/744/1/74}, \href
  {http://adsabs.harvard.edu/abs/2012ApJ...744...74G} {744, 74}

\bibitem[\protect\citeauthoryear{{Gualandris}, {Khan}, {Bortolas}, {Bonetti},
  {Sesana}, {Berczik}  \& {Holley-Bockelmann}}{{Gualandris}
  et~al.}{2022}]{Gualandris2022}
{Gualandris} A.,  {Khan} F.~M.,  {Bortolas} E.,  {Bonetti} M.,  {Sesana} A.,
  {Berczik} P.,   {Holley-Bockelmann} K.,  2022, \mn@doi [\mnras]
  {10.1093/mnras/stac241}, \href
  {https://ui.adsabs.harvard.edu/abs/2022MNRAS.511.4753G} {511, 4753}

\bibitem[\protect\citeauthoryear{{G{\"u}ltekin} et~al.,}{{G{\"u}ltekin}
  et~al.}{2009}]{msigma2009}
{G{\"u}ltekin} K.,  et~al., 2009, \mn@doi [\apj] {10.1088/0004-637X/698/1/198},
  \href {https://ui.adsabs.harvard.edu/abs/2009ApJ...698..198G} {698, 198}

\bibitem[\protect\citeauthoryear{{Izquierdo-Villalba}, {Sesana}  \&
  {Colpi}}{{Izquierdo-Villalba} et~al.}{2023}]{IzquierdoVillalba2023}
{Izquierdo-Villalba} D.,  {Sesana} A.,   {Colpi} M.,  2023, \mn@doi [\mnras]
  {10.1093/mnras/stac3677}, \href
  {https://ui.adsabs.harvard.edu/abs/2023MNRAS.519.2083I} {519, 2083}

\bibitem[\protect\citeauthoryear{{Jiang}, {Cole}, {Sawala}  \& {Frenk}}{{Jiang}
  et~al.}{2015}]{Jiang2015}
{Jiang} L.,  {Cole} S.,  {Sawala} T.,   {Frenk} C.~S.,  2015, \mn@doi [\mnras]
  {10.1093/mnras/stv053}, \href
  {https://ui.adsabs.harvard.edu/abs/2015MNRAS.448.1674J} {448, 1674}

\bibitem[\protect\citeauthoryear{{Khan}, {Berczik}  \& {Just}}{{Khan}
  et~al.}{2018}]{Khan2018}
{Khan} F.~M.,  {Berczik} P.,   {Just} A.,  2018, \mn@doi [\aap]
  {10.1051/0004-6361/201730489}, \href
  {https://ui.adsabs.harvard.edu/abs/2018A&A...615A..71K} {615, A71}

\bibitem[\protect\citeauthoryear{{Khan}, {Mirza}  \&
  {Holley-Bockelmann}}{{Khan} et~al.}{2020}]{Khan2020}
{Khan} F.~M.,  {Mirza} M.~A.,   {Holley-Bockelmann} K.,  2020, \mn@doi [\mnras]
  {10.1093/mnras/stz3360}, \href
  {https://ui.adsabs.harvard.edu/abs/2020MNRAS.492..256K} {492, 256}

\bibitem[\protect\citeauthoryear{{Khochfar} \& {Burkert}}{{Khochfar} \&
  {Burkert}}{2006}]{KhochfarBurkert2006}
{Khochfar} S.,  {Burkert} A.,  2006, \mn@doi [\aap]
  {10.1051/0004-6361:20053241}, \href
  {https://ui.adsabs.harvard.edu/abs/2006A&A...445..403K} {445, 403}

\bibitem[\protect\citeauthoryear{{Kormendy} \& {Ho}}{{Kormendy} \&
  {Ho}}{2013}]{KormendyHo2013}
{Kormendy} J.,  {Ho} L.~C.,  2013, \mn@doi [\araa]
  {10.1146/annurev-astro-082708-101811}, \href
  {https://ui.adsabs.harvard.edu/abs/2013ARA&A..51..511K} {51, 511}

\bibitem[\protect\citeauthoryear{{Mannerkoski}, {Johansson}, {Pihajoki},
  {Rantala}  \& {Naab}}{{Mannerkoski} et~al.}{2019}]{Mannerkoski2019}
{Mannerkoski} M.,  {Johansson} P.~H.,  {Pihajoki} P.,  {Rantala} A.,   {Naab}
  T.,  2019, \mn@doi [\apj] {10.3847/1538-4357/ab52f9}, \href
  {https://ui.adsabs.harvard.edu/abs/2019ApJ...887...35M} {887, 35}

\bibitem[\protect\citeauthoryear{{Marconi} \& {Hunt}}{{Marconi} \&
  {Hunt}}{2003}]{MarconiHunt2003}
{Marconi} A.,  {Hunt} L.~K.,  2003, \mn@doi [\apjl] {10.1086/375804}, \href
  {https://ui.adsabs.harvard.edu/abs/2003ApJ...589L..21M} {589, L21}

\bibitem[\protect\citeauthoryear{{Merritt}}{{Merritt}}{2006}]{Merritt2006}
{Merritt} D.,  2006, \mn@doi [\apj] {10.1086/506139}, \href
  {https://ui.adsabs.harvard.edu/abs/2006ApJ...648..976M} {648, 976}

\bibitem[\protect\citeauthoryear{{Mikkola} \& {Valtonen}}{{Mikkola} \&
  {Valtonen}}{1992}]{MV1992}
{Mikkola} S.,  {Valtonen} M.~J.,  1992, \mn@doi [\mnras]
  {10.1093/mnras/259.1.115}, \href
  {https://ui.adsabs.harvard.edu/abs/1992MNRAS.259..115M} {259, 115}

\bibitem[\protect\citeauthoryear{{Nasim}, {Gualandris}, {Read}, {Dehnen},
  {Delorme}  \& {Antonini}}{{Nasim} et~al.}{2020}]{Nasim2020}
{Nasim} I.,  {Gualandris} A.,  {Read} J.,  {Dehnen} W.,  {Delorme} M.,
  {Antonini} F.,  2020, \mn@doi [\mnras] {10.1093/mnras/staa1896}, \href
  {https://ui.adsabs.harvard.edu/abs/2020MNRAS.497..739N} {497, 739}

\bibitem[\protect\citeauthoryear{{Peters}}{{Peters}}{1964}]{peters1964}
{Peters} P.~C.,  1964, \mn@doi [Physical Review] {10.1103/PhysRev.136.B1224},
  \href {https://ui.adsabs.harvard.edu/abs/1964PhRv..136.1224P} {136, 1224}

\bibitem[\protect\citeauthoryear{{Pillepich} et~al.,}{{Pillepich}
  et~al.}{2018}]{Pillepich2018}
{Pillepich} A.,  et~al., 2018, \mn@doi [\mnras] {10.1093/mnras/stx3112}, \href
  {https://ui.adsabs.harvard.edu/abs/2018MNRAS.475..648P} {475, 648}

\bibitem[\protect\citeauthoryear{{Quinlan}}{{Quinlan}}{1996}]{Quinlan1996}
{Quinlan} G.~D.,  1996, \mn@doi [\na] {10.1016/S1384-1076(96)00003-6}, \href
  {https://ui.adsabs.harvard.edu/abs/1996NewA....1...35Q} {1, 35}

\bibitem[\protect\citeauthoryear{{Rantala}, {Pihajoki}, {Johansson}, {Naab},
  {Lah{\'e}n}  \& {Sawala}}{{Rantala} et~al.}{2017}]{Rantala2017}
{Rantala} A.,  {Pihajoki} P.,  {Johansson} P.~H.,  {Naab} T.,  {Lah{\'e}n} N.,
   {Sawala} T.,  2017, \mn@doi [\apj] {10.3847/1538-4357/aa6d65}, \href
  {https://ui.adsabs.harvard.edu/abs/2017ApJ...840...53R} {840, 53}

\bibitem[\protect\citeauthoryear{{Rantala}, {Johansson}, {Naab}, {Thomas}  \&
  {Frigo}}{{Rantala} et~al.}{2019}]{Rantala2019}
{Rantala} A.,  {Johansson} P.~H.,  {Naab} T.,  {Thomas} J.,   {Frigo} M.,
  2019, \mn@doi [\apjl] {10.3847/2041-8213/ab04b1}, \href
  {https://ui.adsabs.harvard.edu/abs/2019ApJ...872L..17R} {872, L17}

\bibitem[\protect\citeauthoryear{{Rawlings}, {Mannerkoski}, {Johansson}  \&
  {Naab}}{{Rawlings} et~al.}{2023}]{Rawlings2023}
{Rawlings} A.,  {Mannerkoski} M.,  {Johansson} P.~H.,   {Naab} T.,  2023,
  \mn@doi [\mnras] {10.1093/mnras/stad2891}, \href
  {https://ui.adsabs.harvard.edu/abs/2023MNRAS.526.2688R} {526, 2688}

\bibitem[\protect\citeauthoryear{{Rawlings} et~al.,}{{Rawlings}
  et~al.}{2025}]{Rawlings2025}
{Rawlings} A.,  et~al., 2025, \mn@doi [\mnras] {10.1093/mnras/staf238}, \href
  {https://ui.adsabs.harvard.edu/abs/2025MNRAS.537.3421R} {537, 3421}

\bibitem[\protect\citeauthoryear{{Reardon} et~al.,}{{Reardon}
  et~al.}{2023}]{Parks2023}
{Reardon} D.~J.,  et~al., 2023, \mn@doi [\apjl] {10.3847/2041-8213/acdd02},
  \href {https://ui.adsabs.harvard.edu/abs/2023ApJ...951L...6R} {951, L6}

\bibitem[\protect\citeauthoryear{{Reines} \& {Volonteri}}{{Reines} \&
  {Volonteri}}{2015}]{ReinesVolonteri2015}
{Reines} A.~E.,  {Volonteri} M.,  2015, \mn@doi [\apj]
  {10.1088/0004-637X/813/2/82}, \href
  {https://ui.adsabs.harvard.edu/abs/2015ApJ...813...82R} {813, 82}

\bibitem[\protect\citeauthoryear{{Roos}}{{Roos}}{1981}]{Roos1981}
{Roos} N.,  1981, \aap, \href
  {https://ui.adsabs.harvard.edu/abs/1981A&A...104..218R} {104, 218}

\bibitem[\protect\citeauthoryear{{Sesana} \& {Khan}}{{Sesana} \&
  {Khan}}{2015}]{Sesana2015}
{Sesana} A.,  {Khan} F.~M.,  2015, \mn@doi [\mnras] {10.1093/mnrasl/slv131},
  \href {https://ui.adsabs.harvard.edu/abs/2015MNRAS.454L..66S} {454, L66}

\bibitem[\protect\citeauthoryear{{Sesana}, {Haardt}  \& {Madau}}{{Sesana}
  et~al.}{2006}]{Sesana2006}
{Sesana} A.,  {Haardt} F.,   {Madau} P.,  2006, \mn@doi [\apj]
  {10.1086/507596}, \href
  {https://ui.adsabs.harvard.edu/abs/2006ApJ...651..392S} {651, 392}

\bibitem[\protect\citeauthoryear{{Sesana}, {Haardt}  \& {Madau}}{{Sesana}
  et~al.}{2008}]{Sesana2008}
{Sesana} A.,  {Haardt} F.,   {Madau} P.,  2008, \mn@doi [\apj]
  {10.1086/590651}, \href
  {https://ui.adsabs.harvard.edu/abs/2008ApJ...686..432S} {686, 432}

\bibitem[\protect\citeauthoryear{{Wetzel}}{{Wetzel}}{2011}]{Wetzel2011}
{Wetzel} A.~R.,  2011, \mn@doi [\mnras] {10.1111/j.1365-2966.2010.17877.x},
  \href {https://ui.adsabs.harvard.edu/abs/2011MNRAS.412...49W} {412, 49}

\bibitem[\protect\citeauthoryear{{Xu} et~al.,}{{Xu} et~al.}{2023}]{Xu2023}
{Xu} H.,  et~al., 2023, \mn@doi [Research in Astronomy and Astrophysics]
  {10.1088/1674-4527/acdfa5}, \href
  {https://ui.adsabs.harvard.edu/abs/2023RAA....23g5024X} {23, 075024}

\makeatother
\end{thebibliography}







\bsp	
\label{lastpage}
\end{document}